\documentclass[onecolumn,aps,prd,preprintnumbers,showpacs,superscriptaddress,nofootinbib,amsmath,amssymb,floats,floatfix,showkeys,notitlepage]{revtex4-2}

\usepackage{mwe,tikz}
\usepackage{orcidlink}
\usepackage{lipsum}
\usepackage{graphicx}
\usepackage{subfigure}
\usepackage{sans}
\usepackage{changes}
\usepackage{hyperref}
\hypersetup{colorlinks=true,linkcolor=blue,urlcolor=blue,citecolor=blue}
\usepackage[toc,page]{appendix}
\usepackage[normalem]{ulem}
\usepackage{adjustbox}
\usepackage{latexsym}
\usepackage{amsmath}
\usepackage{amssymb}
\usepackage{amsfonts}
\usepackage{dcolumn}
\usepackage{bm}
\usepackage{tikz}
\usepackage{bigints}
\usepackage{array,tabularx,multirow}
\usepackage[tracking=true]{microtype}
\usepackage{mathrsfs}
\usepackage{mathtools}
\usepackage{braket}
\usepackage[utf8]{inputenc}
\SetTracking{}{500}
\SetTracking{encoding={*}, shape=sc}{40}
\UseRawInputEncoding 
\allowdisplaybreaks

\newcommand{\be}{\begin{eqnarray}}
\newcommand{\ee}{\end{eqnarray}}

\def\be{\begin{equation}}
\def\ee{\end{equation}}
\def\bestar{\begin{equation*}}
\def\eestar{\end{equation*}}

\newcommand{\bea}{\begin{eqnarray}}\newcommand{\eea}{\end{eqnarray}}
\newcommand{\brr}{\begin{array}}\newcommand{\err}{\end{array}}
\newcommand{\bit}{\begin{itemize}}\newcommand{\eit}{\end{itemize}}
\newcommand{\ben}{\begin{enumerate}}\newcommand{\een}{\end{enumerate}}

\newcommand{\ba}{\begin{array}}
\newcommand{\ea}{\end{array}}

\graphicspath{{./images/} }

\newcolumntype{M}[1]{>{\centering\arraybackslash}m{#1}}
\newcolumntype{N}{@{}m{0pt}@{}}

\newcounter{sxn}

\newcounter{axn}

\newdimen\mybaselineskip
\newcommand{\beeq}{\begin{equation}}
\newcommand{\eneq}{\end{equation}}
\newcommand{\beqn}{\begin{eqnarray}}
\newcommand{\eeqn}{\end{eqnarray}}

\newcommand{\beal}{\setcounter{letter}{1} \begin{eqnarray}}
\newcommand{\eeal}{\addtocounter{equation}{1} \end{eqnarray}}

\newcommand{\larrow}{\,\,\,\,\hbox to 30pt{\rightarrowfill}
\,\,\,\,}
\newcommand{\slarrow}{\,\,\,\hbox to 20pt{\rightarrowfill}
\,\,\,}

\def\la{\raise.16ex\hbox{$\langle$}\lower.16ex\hbox{}  }
\def\ra{\, \raise.16ex\hbox{$\rangle$}\lower.16ex\hbox{} }

\def\psibar{ \psi \kern-.65em\raise.6em\hbox{$-$} \lower.6em\hbox{} }
\def\psibarb{ \psi \kern-.65em\raise.6em\hbox{$-$}  }

\begin{document} 

\title{Shadow, lensing, quasinormal modes, greybody bounds and neutrino propagation  by dyonic ModMax black holes 
}

\author{Reggie C. Pantig}
\email{rcpantig@mapua.edu.ph}
\affiliation{Physics Department, Map\'ua University, 658 Muralla St., Intramuros, Manila 1002, Philippines}

\author{Leonardo Mastrototaro}
\email{lmastrototaro@unisa.it}
\affiliation{Dipartimento di Fisica ``E.R Caianiello'', Università degli Studi di Salerno, Via Giovanni Paolo II, 132 - 84084 Fisciano (SA), Italy.}
\affiliation{Istituto Nazionale di Fisica Nucleare - Gruppo Collegato di Salerno - Sezione di Napoli, Via Giovanni Paolo II, 132 - 84084 Fisciano (SA), Italy.}

\author{Gaetano Lambiase}
\email{lambiase@sa.infn.it}
\affiliation{Dipartimento di Fisica ``E.R Caianiello'', Università degli Studi di Salerno, Via Giovanni Paolo II, 132 - 84084 Fisciano (SA), Italy.}
\affiliation{Istituto Nazionale di Fisica Nucleare - Gruppo Collegato di Salerno - Sezione di Napoli, Via Giovanni Paolo II, 132 - 84084 Fisciano (SA), Italy.}

\author{Ali \"Ovg\"un}
\email{ali.ovgun@emu.edu.tr}
\affiliation{Physics Department, Eastern Mediterranean University, Famagusta, 99628 North
Cyprus via Mersin 10, Turkey.}

\begin{abstract}
Motivated by recent work on the Modified Maxwell (ModMax) black holes [Phys.Lett.B 10.1016/j.physletb.2020.136011], which are invariant in duality rotations and conformal transformations founded in [ Phys.Rev.D 10.1103/PhysRevD.102.121703], we probe its effects on the shadow cast, weak field gravitational lensing, and neutrino propagation in its vicinity. Using the EHT data for the shadow diameter of Sgr. A* and M87*, and LIGO/VIRGO experiments for the dyonic ModMax black hole perturbations, we find constraints for ModMax parameters such as $Q_\text{m}$ and the screening factor $\gamma$. We also analyze how the shadow radius behaves as perceived by a static observer and one that is comoving with the cosmic expansion. The effect of the ModMax parameters is constant for a static observer, and we found That it varies when the observer is comoving with cosmic expansion. We also analyzed its effect on the weak deflection angle by exploiting the Gauss-Bonnet theorem and its application to Einstein ring formation. We also consider the finite distance effect and massive particle deflection. Our results indicate that the far approximation of massive particle gives the largest deflection angle and amplifies the effect of $Q_\text{m}$ and $\gamma$. Then we also calculate the quasinormal modes and greybody bounds which encode unique characteristic features of the dyonic ModMax black hole. With the advent of improving space technology, we reported that it is possible to detect the deviation caused through the shadow cast, Einstein rings, quasinormal modes, and neutrino oscillations.
\end{abstract}

\date{\today}
\keywords{Black hole; ModMax non-linear electrodynamics; Shadow; Quasinormal modes; Greybody; Neutrino oscillation}

\pacs{95.30.Sf, 04.70.-s, 97.60.Lf, 04.50.+h}

\maketitle

\tableofcontents

\section{Introduction} \label{sec1}
One of the most important problems in Einstein's theory of general relativity is the singularities at the beginning of the universe and black hole solutions. There are similar singularities in Maxwell's theory of  Electrodynamics \cite{Gibbons:2001gy}. To avoid these singularities, firstly, Born-Infeld (BI) modified the Maxwell theory of electrodynamics in 1934, which is also relativistic and gauge-invariant theory, known as BI nonlinear electrodynamics (NED) \cite{Born:1934gh}. In BI NED, the self-energy of charges is finite. Moreover, the effective action of BI NED can also be derived from open superstrings at low energy dynamics of D-branes, without any physical singularities, \cite{Fradkin:1985qd,Gibbons:2001gy}. Another example is Euler-Heisenberg (EH) NED, which occurs due to the polarization of the vacuum \cite{Heisenberg:1936nmg}. Both BI NED and EH NED (which are SO(2) electric-magnetic duality invariant) reduce to the Maxwell electrodynamics in the weak field regime because of the fixed energy scale interactions which break the conformal invariance. Afterwards, based on the Einstein-NED theories, the regular black hole solutions were obtained, with freedom of duality rotations \cite{Ayon-Beato:1998hmi,Bronnikov:2000yz,Bronnikov:2000vy}. Recently authors in \cite{Bandos:2020jsw} constructed a generalization of Maxwell electrodynamics known as ModMax electrodynamics which has a low-energy limit of a one dimensionless parameter generalization of BI, and $\gamma=0$ condition gives Maxwell's equations. Recently, Flores-Alfonso et al. have found new black hole solutions in ModMax electrodynamics \cite{Flores-Alfonso:2020euz}. The SO(2) invariance for electric and magnetic fields gives us the dyonic solutions. The effect of ModMax electrodynamics on black hole spacetimes via screening factor $\gamma$ shields the actual charges. Then ModMax electrodynamics have been considered by many authors  \cite{Banerjee:2022sza,Bokulic:2022cyk,Lechner:2022qhb,Barrientos:2022bzm,Ortaggio:2022ixh,Nastase:2022fzx,Ferko:2022iru,Ali:2022yys,Kruglov:2022qag,Sorokin:2021tge,Escobar:2021mpx,Nastase:2021uvc,Bokulic:2021xom,Nomura:2021efi,Zhang:2021qga,Kruglov:2021bhs,Bandos:2021rqy,Bokulic:2021dtz,Flores-Alfonso:2020nnd,BallonBordo:2020jtw,Babaei-Aghbolagh:2022uij}. 

One of the most striking features of a compact object such as a black hole is the gravitational bending of light in its vicinity. It introduces the innermost region for photon orbit, which is unstable, and any perturbation will cause photons to spiral into the black hole or escape to infinity. These photons that escape to infinity determine the shadow cast perceived by an observer at infinity. The Event Horizon Telescope Collaboration's \cite{EventHorizonTelescope:2019dse,EventHorizonTelescope:2022xnr}   experimental findings revealed the first image of the black holes M87* and Sgr. A*, which demonstrates the existence of a black hole shadow that was theoretically studied by \cite{Synge:1966} and \cite{Luminet:1979}. While the literature on the study of shadow cast is quite exhaustive before and after the release of the black hole image by the EHT, it leaves no doubt about the importance of studying the black hole shadow since it reveals imprints of the spacetime being considered, and the effects of any type of astrophysical environments \cite{Allahyari:2019jqz,Vagnozzi:2020quf,Vagnozzi:2019apd,Khodadi:2021gbc,Ovgun:2018tua,Ovgun:2020gjz,Pantig:2022toh,Uniyal:2022vdu,Pantig:2022qak,Lobos:2022jsz,Rayimbaev:2022hca,Pantig:2022ely,Pantig:2022whj,Pantig:2022sjb,Afrin:2021imp,Kumar:2020hgm,Kumar:2018ple,Kumar:2020owy,Atamurotov:2013sca,Belhaj:2020rdb,Gralla:2019xty,Papnoi:2021rvw,Atamurotov:2015nra,Atamurotov:2015xfa,Papnoi:2014aaa}. It led to various methods being developed to explore such effects \cite{Tsupko2020,Perlick_2015,Perlick:2018,Roy:2020dyy,Chang:2020miq,Dokuchaev:2019jqq,Dokuchaev:2020wqk,Dokuchaev:2019bbf,Dokuchaev:2018ibr}. This study aims to constrain the screening factor $\gamma$ using the EHT data and explore how the shadow cast/radius behaves. We will also analyze a more realistic scenario, the shadow cast as perceived by a non-static observer, that co-moves with the cosmic expansion. Finally, we extend the situation to different cosmological models.

According to Einstein's theory of general relativity, masses deflect light in a way similar to convex glass lenses known as gravitational lensing, which is a powerful tool to test alternative gravity theories \cite{Virbhadra:1999nm,Virbhadra:2002ju,Virbhadra:1998dy,Virbhadra:2007kw,Virbhadra:2008ws,Adler:2022qtb}. In astrophysics, distances play a dominant role in determining the properties of astrophysical objects. Contrary to this, Virbhadra showed that just observation of relativistic images  (no information about the masses and distances are required) gives an incredibly accurate value for the upper bound to the compactness of massive dark objects \cite{Virbhadra:2022ybp}. Moreover, Virbhadra hypothesized that there exists a distortion parameter such that the signed sum of all images of singular gravitational lensing of a source identically vanishes by testing this with images of Schwarzschild lensing in weak and strong gravitational fields \cite{Virbhadra:2022iiy}.  

Although there are many methods to calculate weak deflection angle, Gibbons and Werner proposed an alternative method to calculate it by applying the Gauss-Bonnet theorem (GBT) on the optical metric of the asymptotically flat black holes \cite{Gibbons:2008rj}. Werner then improved this method for rotating spacetimes such as Kerr black hole \cite{Werner:2012rc}. Afterwards, this method was applied to various asymptotically flat spacetimes and was also extended to non-asymptotically flat spacetimes \cite{Ovgun:2018fnk,Ovgun:2019wej,Li:2020dln,Ovgun:2018oxk,Javed:2019rrg,Javed:2019ynm,Li:2020wvn,Kumaran:2019qqp,Javed:2020lsg,Ovgun:2020gjz,Ovgun:2020yuv,Pantig:2022toh,Pantig:2022sjb,Pantig:2022whj,Uniyal:2022vdu,Pantig:2022qak,Pantig:2022ely,Rayimbaev:2022hca,Lobos:2022jsz,Javed:2019qyg,Jusufi:2017mav,Ovgun:2018tua,Jusufi:2017lsl,Sakalli:2017ewb}.

Detection of the gravitational waves by LIGO/VIRGO collaborations opens a new gate in black hole physics. The gravitational waves provide the characteristic vibration modes of black holes (complex frequency of damped oscillations) or imprint of the black holes, independent of what exactly excited the modes), which are known as quasinormal modes (QNMs). In the 1950s, the pioneering work on black hole perturbations was done by Regge and Wheeler \cite{Regge:1957td} and then by Zerilli \cite{Zerilli:1970wzz}. In the 1960s, Thorne studied the perturbations of relativistic stars in GR \cite{Thorne:1968zz,Thorne:1997cw}. First, Vishveshware imagined and studied the QNMs of the Schwarzschild black hole by calculating the scattering of gravitational waves \cite{Vishveshwara:1970zz}. Then Press used the term of quasinormal frequencies (QNF) \cite{Press:1971wr}. Afterwards, Davis et al. showed how to obtain QNM oscillations from the infalling particle into Schwarzschild black hole \cite{Davis:1971gg} and Detweiler et al. showed for Kerr black holes \cite{Detweiler:1980uk}. Black hole perturbations and QNMs are a well-studied and active field to study, and there are various methods on different spacetimes in literature (for more information \cite{Daghigh:2011ty,Daghigh:2020mog,Daghigh:2008jz,Zhidenko:2003wq,Zhidenko:2005mv,Chabab:2017knz,Lepe:2004kv,Yang:2022ifo,Yang:2022xxh,Liu:2022ygf,Fernando:2022wlm,Fernando:2012yw,Fernando:2003ai,Fernando:2008hb,Liu:2021xfb,Yang:2021cvh}, see these reviews \cite{Kokkotas:1999bd,Nollert:1999ji,Berti:2009kk}). The QNMs could be detected through the gravitational wave interferometers, so we would like to explore the properties of background spacetime of ModMax black holes and try to probe the ModMax BH parameters by studying gravitational waves (GWs) at the ringdown stage.

In 1974 Hawking showed that there are 'grey holes' and not black holes because of their radiation, which is impossible classically for ingoing particles \cite{Hawking:1975vcx}. The quantum field theory calculations near the black hole's horizon provide the emission of quantum radiation due to the creation and annihilation of particles. The Hawking radiation is moving on a curved spacetime ( as a potential barrier) so that some radiations are reflected into the black hole and the rest travel to spatial infinity \cite{Akhmedov:2006pg}. These deviations from the blackbody radiation spectrum, as seen by an asymptotic observer, are known as the grey-body factor, which is a synonym for transmission probability \cite{Maldacena:1996ix,Harmark:2007jy,Rincon:2018ktz,Rincon:2020cos,Panotopoulos:2018pvu,Panotopoulos:2017yoe,Klebanov:1997cx,Liu:2022ygf}, and there are various methods to calculate it. One can use the matching technique \cite{Fernando:2004ay}, the WKB approximation \cite{Konoplya:2019hlu} or the rigorous bound \cite{Visser:1998ke,Boonserm:2008zg}. On the other hand, in 1968, Matzner studied the absorption and scattering of a massive scalar field from hitting a black hole \cite{Matzner}; hence he showed that the total cross-section for the absorption process vanishes. In 1978 Sanchez calculated the absorption spectrum of the Schwarzschild black hole and obtained the total absorption cross section in the Hawking formula \cite{Sanchez:1977si}. Sanchez showed that the absorption spectrum as a function of the frequency makes clear oscillations characteristic of a diffraction pattern, where the oscillations occur around the constant value of the geometrical optics with decreasing amplitude and constant period. The absorption cross-section for black holes has been studied in the literature for various black hole spacetimes \cite{Fabbri:1975sa,Unruh:1976fm,Andersson:1995vi,Benone:2014qaa,Macedo:2014uga,Crispino:2015gua,Leite:2017zyb,Javed:2022vpy,Javed:2022rrs,Javed:2022kzf,Yang:2022ifo,Javed:2021ymu,Okyay:2021nnh,Pantig:2022ely}, especially for the low-frequency behaviour of the absorption and for the zero-frequency limit (it becomes equal to the surface area of the black hole horizon) \cite{Unruh:1976fm,Das:1996we,Higuchi:2001si}, as well as the high-frequency behaviour of the scalar absorption via Sinc approximation \cite{Sanchez:1977si,Decanini:2009mu,Decanini:2011xi}. 


In this paper, we also investigate the propagation of neutrinos in dyonic ModMax BH. We focus, in particular, on the neutrino flavour oscillation and spin-flip of neutrinos when they scatter off BH. Clear evidence that neutrinos may oscillate in different flavours comes from experiments on neutrinos \cite{Ace19,Ker20,Aga19}. These experiments indicate that neutrinos are massive particles and, therefore, that physics beyond the Standard Model is required to incorporate the neutrino particle properly. An important aspect related to the properties of neutrinos is their interaction with external fields. These interactions can be either electromagnetic or gravitational. In these cases, the formulas of the oscillation probabilities of mixed neutrinos (different flavours) are affected by the external field, compared to the formulas computed in vacuum \cite{BalKay18}. For example, the interactions of neutrinos with external electromagnetic fields may induce the so-called spin oscillation and/or spin-flavour oscillations, that is a helicity transition of neutrinos with different helicities \cite{Giu19}. It is also well known the gravitational interaction or propagation of neutrinos in the curved background can affect the neutrino oscillations probability, or induce the change of the polarization of a spinning particle \cite{Pap51,PirRoyWud96,SorZil07,ObuSilTer17,Dvo06,Dvo19,Cuesta,luca,Cardall,Visinelli,Chakraborty,Sor12,Ahluwalia,Punzi,Swami,CuestaApJ,LambMNRAS,Capozz,Lambiase:2021txu,Lambiase:2022ucu,Capolupo:2020wlx}. In this paper we study the neutrino oscillation probability for neutrinos propagating in dyonic ModMax  BH. We also study helicity transitions of neutrinos, hence the transition $\nu_{fL}\to\nu_{fR}$, where $f=e,\mu\tau$, corresponding to the case in which neutrino flavour remains fixed (the generic case of neutrino spin-flavour oscillations has been studied in \cite{Sadjadi:2020ozc}). This analysis is important because neutrinos are produced with a fixed left-handed polarization in the Standard Model so that if a transition to right-handed polarization occurs, these neutrinos become sterile (they interact only gravitationally). As a consequence, a detector would register a different neutrino flux, providing in such a way a signature of the coupling of neutrinos with the gravitational background. Here we shall consider only the gravitational field described by ModMax BH.

Our work's layout is as follows: Sect. \ref{sec2} is devoted to investigating the shadow cast behaviour and the effect of a co-moving observer. In Sect. \ref{sec3}, we analyzed the behaviour of the weak deflection angle of both massive and null particles. Then, in Sect. \ref{sec4} - \ref{sec6}, we study the effect of the screening parameter on the spherical infalling accretion, quasinormal modes, and bounds of the greybody factors. Finally, in Sect. \ref{sec7}, we investigate its effect on neutrino oscillations. In particular, we focus on the neutrino flavour and spin transition in the background described by a dyonic ModMax BH. We shortly analyzed also the effect on nucleosynthesis processes. Conclusions are discussed in Section \ref{sec8}. Throughout this paper, we used geometrized units $G = c = 1$, and the metric signature $(-,+,+,+)$.

\section{Dyonic ModMax black holes}
\begin{figure}
   \centering
    \includegraphics[width=0.48\textwidth]{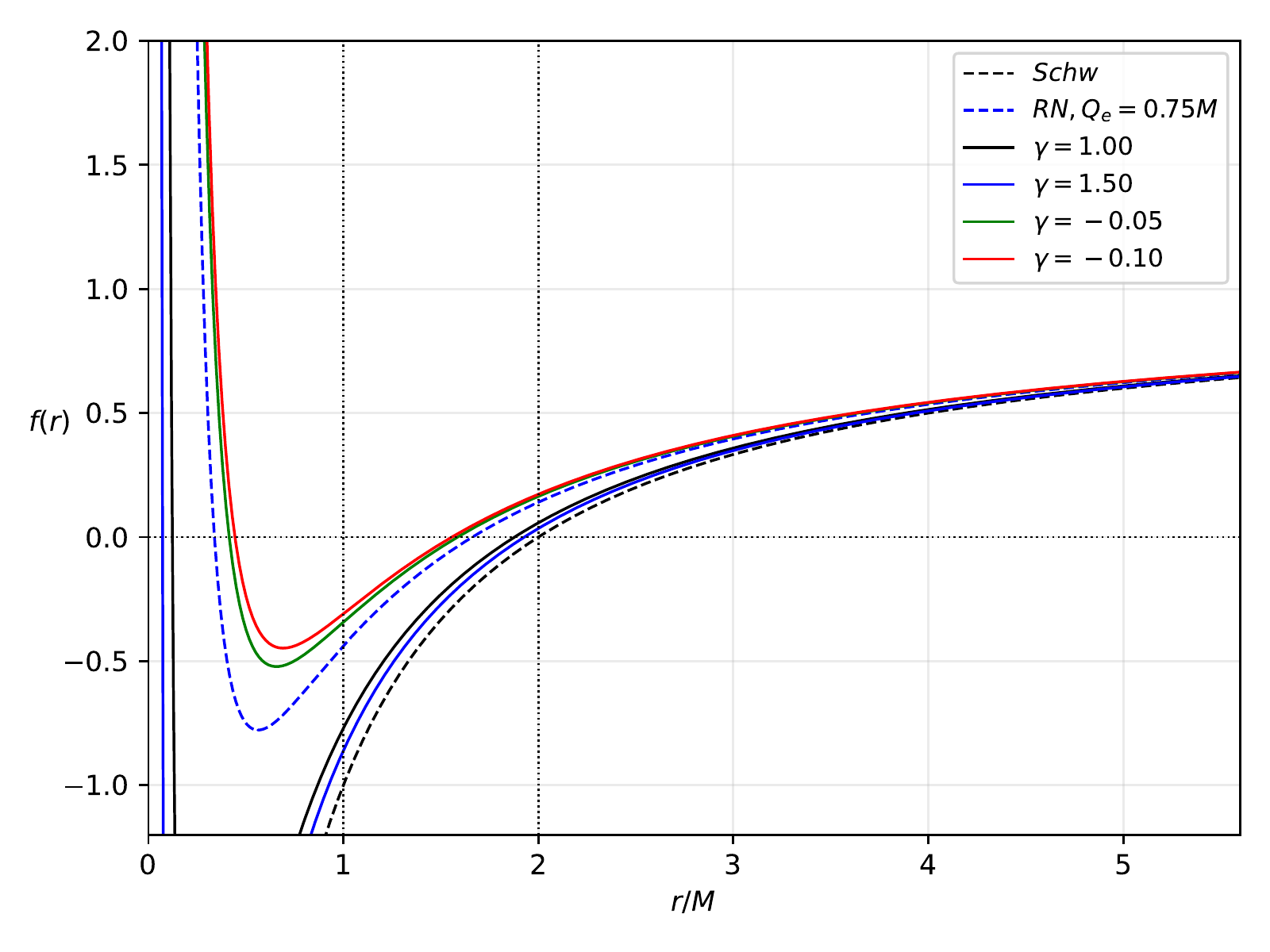}
    \caption{Plot of the $f(r)$, which reveals the location of the inner and outer horizons due to the dyonic ModMax black hole. Here, $Q_\text{m}=0.25M$.} 
    \label{lapse}
\end{figure}

In this section, we study the  spherically symmetric metric of Dyonic ModMax black hole solutions derived in \cite{Flores-Alfonso:2020euz} as follows;
 
\be
ds^2 = -f(r) dt^2 + \frac 1{f(r)} dr^2 + r^2 d\Omega,
\label{ds2}
\ee 
where $d\Omega := d\theta^2 + \sin^2\theta d\phi^2$ and $(t,r,\theta,\phi)$ with the metric function:
\be
f_\text{dyon}(r) = 1 - \frac {2M}r + \frac {(Q_\text{e}^2 + Q_\text{m}^2)e^{-\gamma}}{r^2}.
\label{eq:fdyon}
\ee
We plot the above metric function shown in Fig. \ref{lapse}. Here, we can see theoretically that only a single horizon is formed for the Schwarzschild case, while there are inner and outer horizons for the RN case when $Q_\text{e} = 0.75M$. As the plot indicates, we can observe how the screening parameter affects these null boundaries. The event horizon of the ModMax black hole is located at:
\be
r_\pm = M \pm \sqrt{M^2 - Q_\text{e}^2 e^{-\gamma}}.
\ee 
 Note that the mass must always be larger than the charge; however, it is not valid for the case  of the ModMax black hole because, in the extremal case of the ModMax black hole, it occurs at $r_+ = r_-$:
\be
M_\text{ext} = |Q_\text{e}| e^{-\gamma/2},
\ee
so that  $\gamma > 0$, then $M_\text{ext} < |Q_\text{m}|$; similarly with non-linear electrodynamics~\cite{Ruffini:2013hia,Maceda:2018zim}.

\section{Shadow cast} \label{sec2}
In this section, we first analyze the behaviour of the ModMax black hole's shadow as the screening parameter $\gamma$ varies. Since spacetime has spherical symmetry, we can analyze null geodesics $\theta = \pi/2$, along the equatorial plane. To derive the null particle's equation of motion, consider the Lagrangian
\begin{equation}
    \mathcal{L} = \frac{1}{2}\left( -A(r) \dot{t} +B(r) \dot{r} + C(r) \dot{\phi} \right),
\end{equation}
where we wrote $A(r)=f(r)$, $B(r)=1/f(r)$ and $C(r)=r^2$. With the least action principle, two constants of motion can be derived:
\begin{equation}
    E = A(r)\frac{dt}{d\lambda}, \quad L = C(r)\frac{d\phi}{d\lambda},
\end{equation}
where we can define a useful constant called the impact parameter:
\begin{equation}
    b \equiv \frac{L}{E} = \frac{C(r)}{A(r)}\frac{d\phi}{dt}.
\end{equation}
Going back to the metric, $ds^2=0$ for null geodesics allows us to derive how the $r$ changes with $\phi$. In other words,
\begin{align}
    \left(\frac{dr}{d\phi}\right)^2 =\frac{C(r)}{B(r)}\left(\frac{h(r)^2}{b^2}-1\right),
\end{align}
where we have defined
\begin{equation} \label{eh(r)}
    h(r)^2 = \frac{C(r)}{A(r)}.
\end{equation}
With the above function, one can solve or locate the photonsphere radii via $h'(r)=0$ where the prime denotes the derivative with respect to $r$. For our case, we obtained
\begin{equation} \label{erph}
    r_\text{ph}=\frac{3M}{2}\pm\frac{1}{2}\sqrt{9M^{2}-8\left(Q_\text{e}^2 + Q_\text{m}^2\right)e^{-\gamma}}.
\end{equation}
A non-rotating black hole, such as in this study, only produces a circular shadow silhouette. The shadow is caused by photons escaping the photonsphere due to tiny perturbations. In the Schwarzschild case, in particular, $r_\text{ph} = 3M$, while the shadow radius is $R_\text{sh} = 3\sqrt{3}M$. It is important to note that this shadow radius coincides with the critical impact parameter $b_\text{crit}$. In some situations, if the cosmological constant is considered, this is not the case since the photons are affected by the astrophysical environment as it travels toward the receiver, usually a remote and static observer \cite{Perlick:2018}.

Considering the static observer and following the formalism found in Ref. \cite{Perlick:2018}, a careful inspection of the line element and situation allows us to define
\begin{equation}
    \tan(\alpha_{\text{sh}}) = \lim_{\Delta x \to 0}\frac{\Delta y}{\Delta x} = \left(\frac{C(r)}{B(r)}\right)^{1/2} \frac{d\phi}{dr} \bigg|_{r=r_\text{obs}},
\end{equation}
or, written in another way,
\begin{equation}
    \cot^2(\alpha_{\text{sh}}) = \left(\frac{B(r)}{C(r)}\right) \left. \left(\frac{dr}{d\phi}\right)^2 \right|_{r=r_\text{obs}}.
\end{equation}
With the orbit equation and elementary trigonometry, it is easy to see that
\begin{equation} \label{eangrad}
    \sin^{2}(\alpha_\text{sh}) = \frac{b_\text{crit}^{2}}{h(r_\text{obs})^{2}},
\end{equation}
where $ b_\text{crit} $ is associated with the photonsphere radius. It can be derived by satisfying the condition $d^2r/d\phi^2=0$ and imposing $r \to r_\text{ph}$:
\begin{equation} \label{ebcrit}
    b_\text{crit}^2 = \frac{h(r)}{\left[B'(r)C(r)-B(r)C'(r)\right]} \Bigg[h(r)B'(r)C(r)-h(r)B(r)C'(r)-2h'(r)B(r)C(r) \Bigg],
\end{equation}
where in our case yields
\begin{equation} \label{ebcrit2}
    b_\text{crit}^2 = \frac{3r_\text{ph}^3}{r_\text{ph}-M}.
\end{equation}
Using Eqs. \eqref{eh(r)}, \eqref{erph}, and \eqref{ebcrit2} to Eq. \eqref{eangrad}, we obtain an exact formula for the shadow radius as
\begin{equation} \label{eshadexact}
    \mathcal{R}_{\text{sh}} = r_\text{ph}\left[2\left(1-\frac{2M}{r_\text{obs}}+\frac{\left(Q_\text{e}^2 + Q_\text{m}^2\right)e^{-\gamma}}{r_\text{obs}^{2}}\right)\left({1-\frac{M}{r_\text{ph}}}\right)^{-1}\right]^{1/2}.
\end{equation}

Let us first discuss the observational constraint of the screening parameter using the obtained data from M87* \cite{EventHorizonTelescope:2019dse} and Sgr. A* \cite{EventHorizonTelescope:2022xnr}. First and foremost, these astrophysical black holes are indeed rotating. However, as pointed out in Ref. \cite{Vagnozzi:2022moj}, the constraints for the spin parameter $a$ on Sgr. A* is not clear, but recent constraints place the spin parameter at $a \lesssim 0.1M$ due to the impact of frame-dragging precession on the orbit of S-stars \cite{Fragione:2020khu}. Similar uncertainties for the spin of M87 are also present: Using accretion physics, recent estimates have shown that $0.20 < a < 0.50$ \cite{Garofalo:2020ajg}. Other authors \cite{Nemmen:2019idv} reported that through the measurements of the average power of the relativistic jet and using the upper limit of mass accretion rate for black holes in general relativistic magnetohydrodynamic models of jet formation, the spin parameter of M87* would likely fall from $a \geq 0.40M$ for the prograde case, and $a \geq 0.50M$ for the retrograde case. It was also shown in Ref. \cite{Vagnozzi:2022moj} that slow spin does not affect the shadow radius dramatically, and $a \sim 0$ can be safely assumed for constraining parameters coming from different black hole models. Finally, astrophysical black holes such as these are more likely to contain no charge due to the neutralization process of the accreting ionized plasma. In Ref. \cite{Zajacek:2018ycb}, constraints indicate that $Q_\text{m}/M$ can be very close to zero. Nonetheless, in the following analysis, we will assume the role played by $Q_\text{e}$ and $Q_\text{m}$, as well as the parameter $\gamma$. We will analyze the Modmax black hole for low and average values of $Q_\text{e}$. Finally, we need to remark that we will only analyze the \textit{classical} shadow silhouette, where its boundary is dominated by the glowing accretion disk causing it to become invisible. The dark spot in the EHT image of M87* and Sgr. A* is the imprint of the event horizon's shadow. It is caused by highly redshifted escaping photons coming from accreting matter near the event horizon. As a result, this shadow is smaller than the classical shadow \cite{Dokuchaev:2019jqq}.

According to the seminal EHT papers \cite{EventHorizonTelescope:2019dse, EventHorizonTelescope:2022xnr}, the reported classical shadow angular diameter is $\theta_\text{M87*} = 42 \pm 3 \:\mu$as. Other parameters are the distance of the M87* from the Earth, which is $D = 16.8$ Mpc, and the mass of the M87* is $M_\text{M87*} = 6.5 \pm 0.90$x$10^9 \: M_\odot$. For Sgr. A* the angular shadow diameter is $\theta_\text{Sgr. A*} = 48.7 \pm 7 \:\mu$as (EHT), the distance of the Sgr. A* from the Earth is $D = 8277\pm33$ pc and the mass of the black hole is $M_\text{Sgr. A*} = 4.3 \pm 0.013$x$10^6 \: M_\odot$ (VLTI). Let us now calculate the diameter of the shadow size in units of the black hole mass using
\begin{equation} \label{ed}
    d_\text{sh} = \frac{D \theta}{M},
\end{equation}
which is just the standard arc-length formula. Meanwhile, the theoretical shadow diameter can be obtained via $d_\text{sh}^\text{theo} = 2\mathcal{R}_\text{sh}$. Therefore, by using Eq. \eqref{ed}, we get the diameter of the shadow image of M87* and Sgr. A* as $d^\text{M87*}_\text{sh} = (11 \pm 1.5)M$, and $d^\text{Sgr. A*}_\text{sh} = (9.5 \pm 1.4)M$ respectively.
\begin{figure*}
   \centering
    \includegraphics[width=0.48\textwidth]{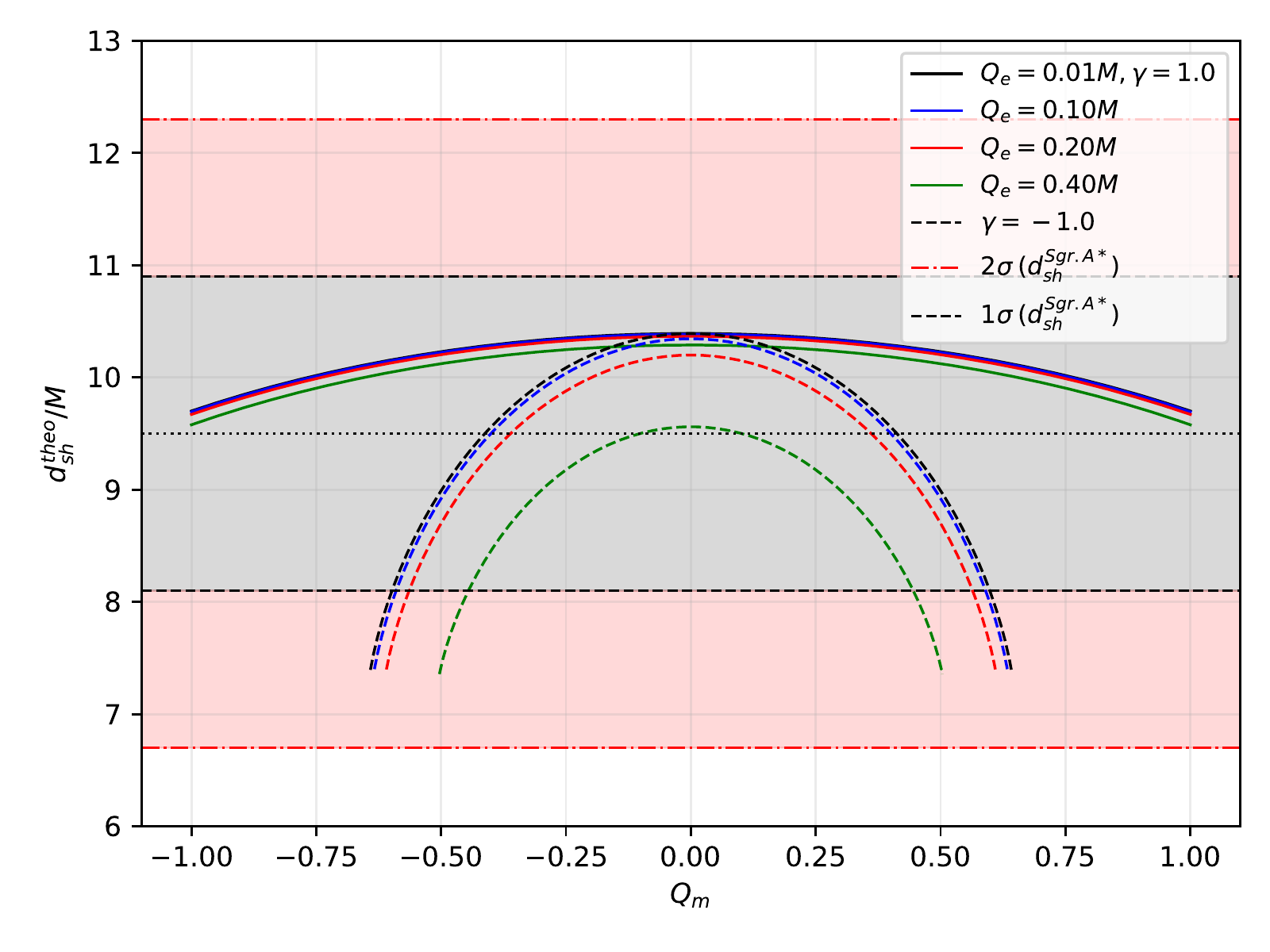}
    \includegraphics[width=0.48\textwidth]{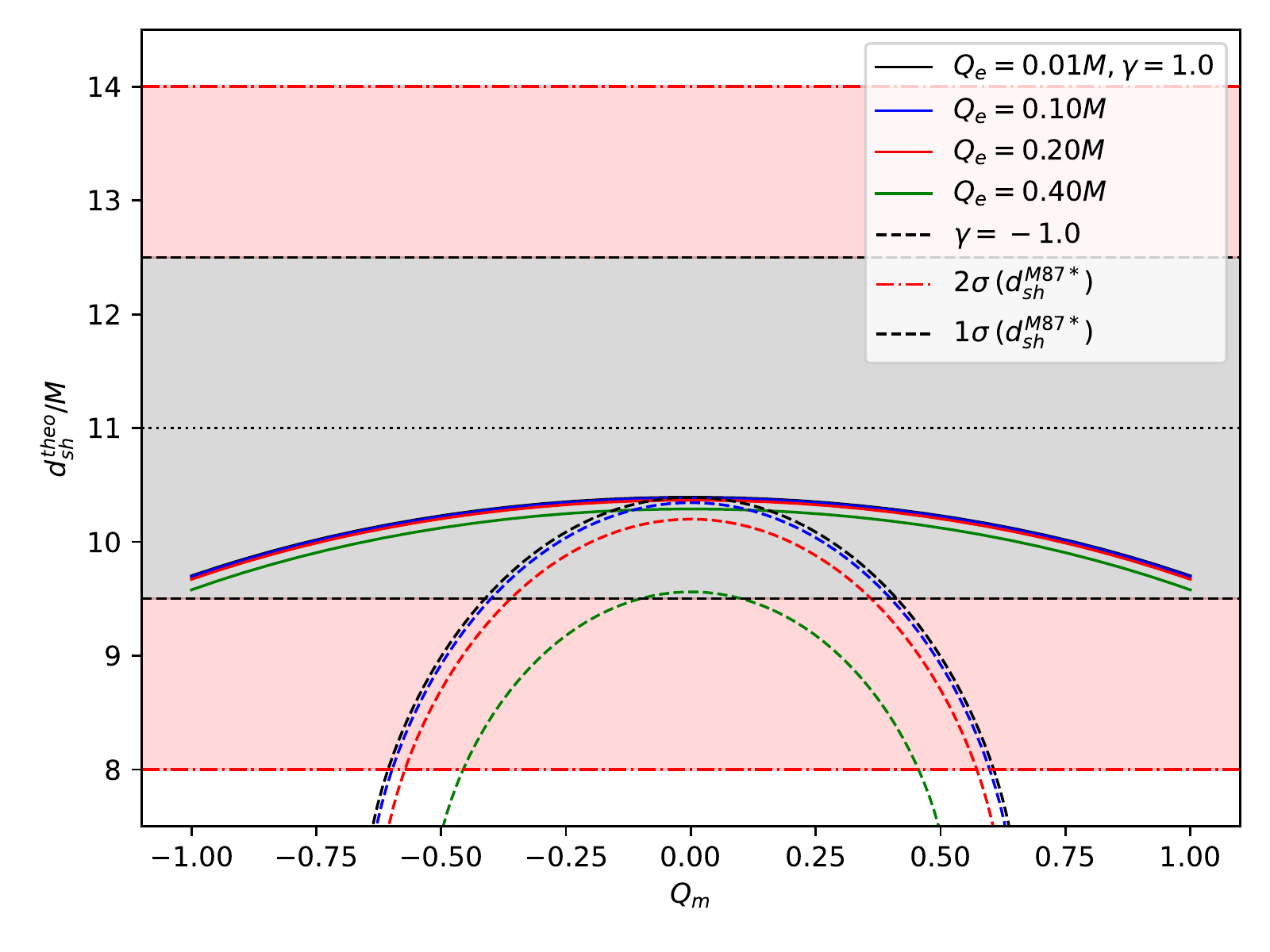}
    \caption{Observational constraints. Left: Sgr. A*, right: M87*. The same plot appearance is obtained for $Q_\text{m} < 0$. }
    \label{sha_cons}
\end{figure*}
In Fig. \ref{sha_cons}, we plotted the allowed values of $Q_\text{m}$ while maintaining the screening parameter $\gamma$ fixed. Such a plot will give how the shadow diameter behaves at our location from Sgr. A* and M87* as we vary $Q_\text{e}$ and $Q_\text{m}$. We included a very small value for $Q_\text{e}$ to represent the notion that these astrophysical black holes may contain a negligible charge. We see that if $\gamma = 1$, it amplifies the effect of $Q_\text{e}$ and $Q_\text{m}$, in contrast with $\gamma = -1$. Nevertheless, we observe how  $\gamma$ affects the curve. An important observation of this plot is that as $\gamma$ decreases, the line becomes more curved, allowing smaller values of $\pm Q_\text{m}$ within the $1\sigma$ uncertainty. It is also worth noting that if $Q_\text{e}$ is close to zero, the value of $\gamma$ is irrelevant when $Q_\text{m}$ is also close to zero. While it happens above the mean for Sgr. A*, it happens below the mean for M87*. As a result, the range for $Q_\text{m}$ at $1\sigma$ level is lesser than Sgr. A*.

\begin{figure*}
   \centering
    \includegraphics[width=0.48\textwidth]{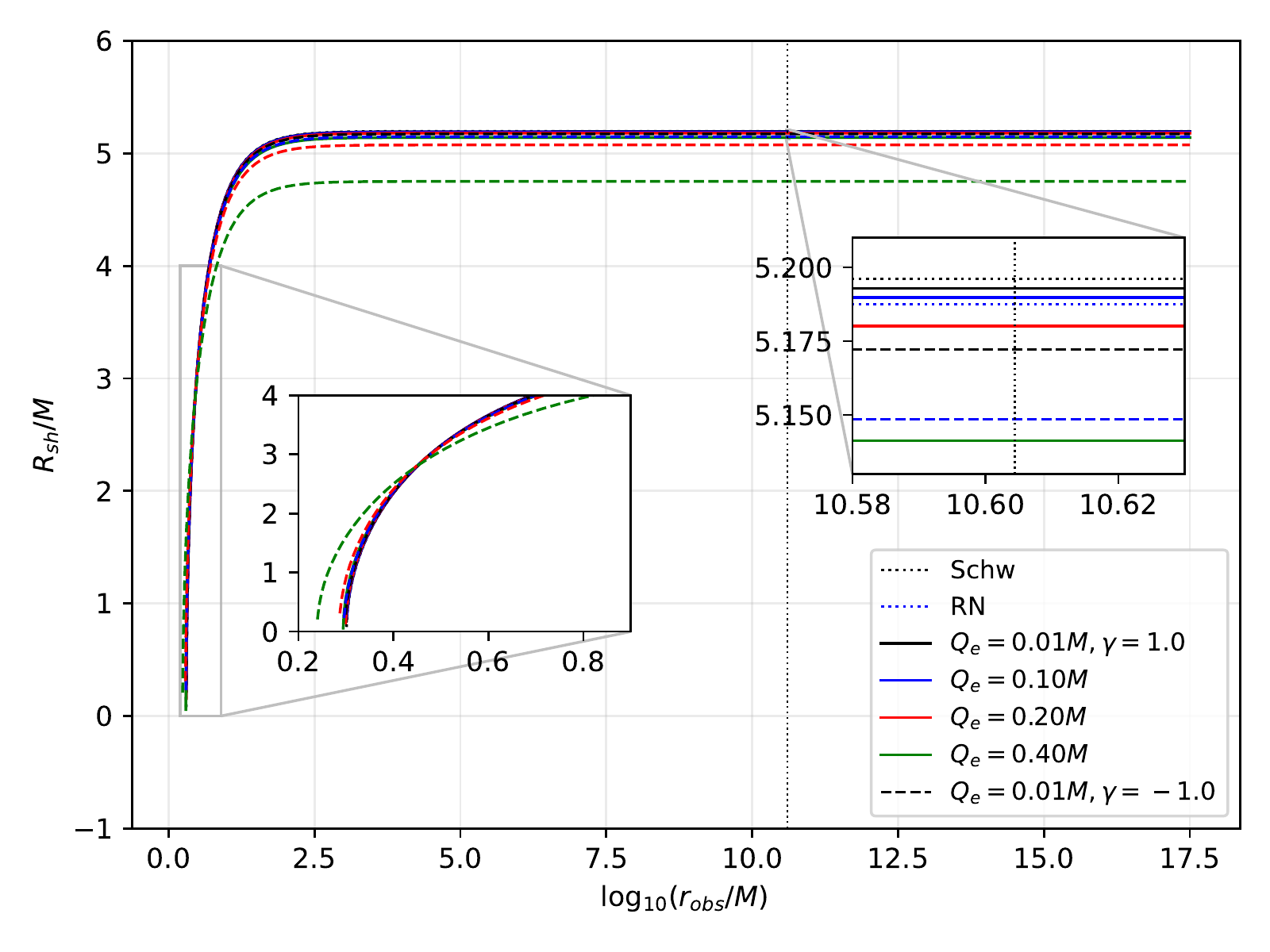}
    \includegraphics[width=0.48\textwidth]{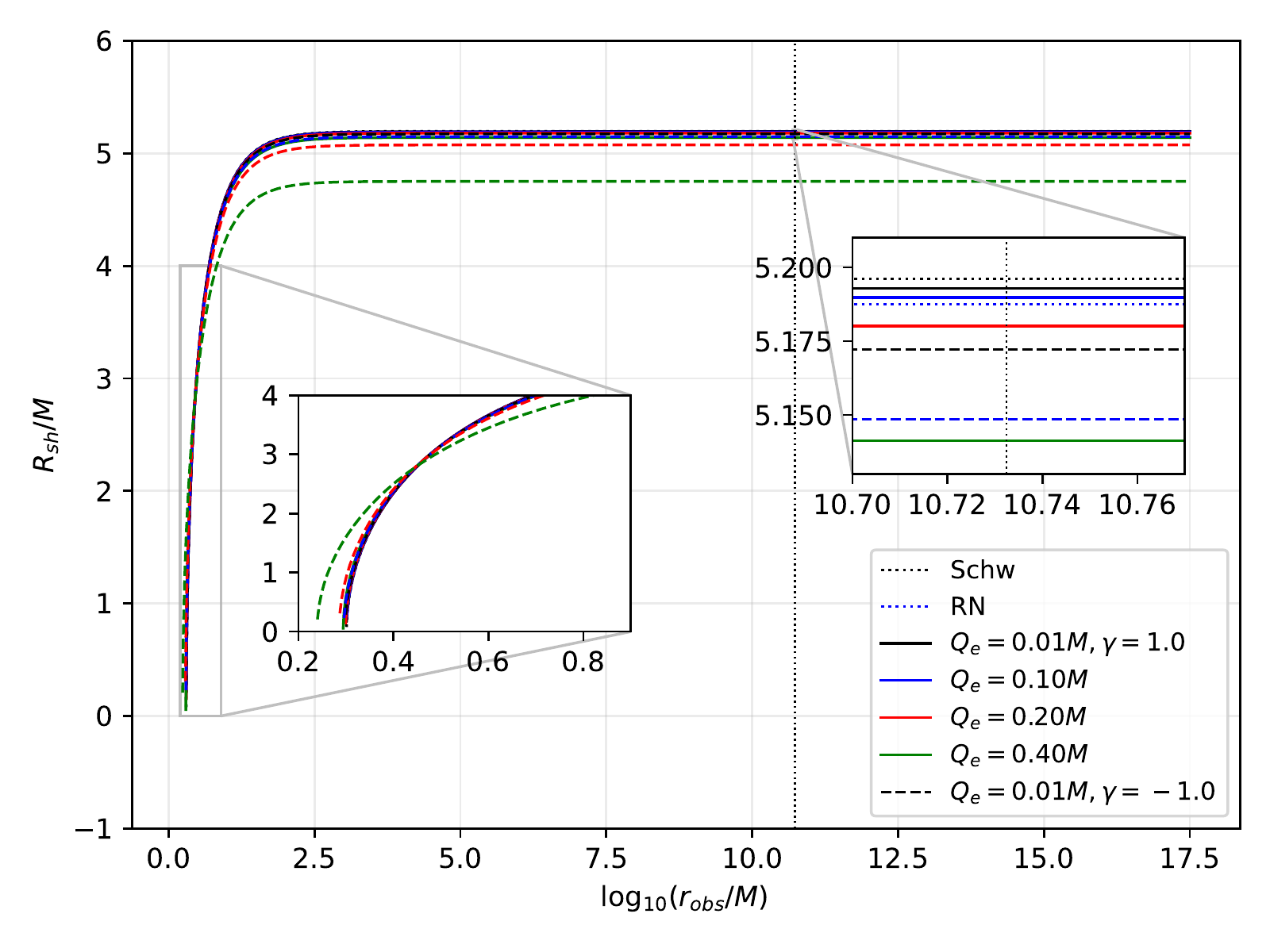}
    \caption{Behavior of shadow radius for a static observer located at $r_\text{obs}$. Left: Sgr. A*. Right: M87*. Here, we have chosen $Q_\text{m} = 0.10 M$, which fits the constrained values in Fig. \ref{sha_cons}. The shadow behaviour for M87* follows the same pattern. }
    \label{sha_exact}
\end{figure*}

Let us now plot how a static observer, at different points $r_\text{obs}$, will perceive the shadow size of Sgr. A* and M87*. We plotted this in Fig. \ref{sha_exact}, along with the Schwarzschild and RN cases, where $Q = 0.10 M$ to see its effect. Our current position from these SMBHs is represented in the right inset plot, where we zoomed in on the effects of the screening parameter $\gamma$ and $Q_\text{m}$, manifest in visuals in Fig. \ref{sha_cons}. The pattern is the same for M87*, except that we are much farther. Such a similarity is expected because the ModMax parameters have no dependence on $r$. Finally, it is interesting how the shadow behaves near the black hole, where we can see an abrupt increase in the shadow radius until $r_\text{obs}/M \sim 316$. The case where $\gamma < 0$ shows that the shadow forms early or nearer the event horizon, compared to the case where $\gamma > 0$. However, as $r_\text{obs}$ increases, $\gamma > 0$ gives a larger shadow. The point where the curves intersect is also interesting, for it represents that $\theta_{sh} = \pi/2$, indicating that in such a location, half of the sky is dark \cite{Perlick:2018}. 

\subsection{Behavior of the shadow due to a co-moving observer}
In reality, observers are not static. Thus, this study aims to investigate the effect of the screening parameter on the shadow perceived by an observer co-moving with the cosmological expansion of the Universe. Here, we apply the recent formalism in Ref. \cite{Tsupko2020}, first focusing on the dark energy-dominated Universe. Then, we also extend the analysis to matter and radiation-dominated Universes.

An observer in a co-moving frame is better understood using the McVittie metric. According to Ref. \cite{Gao2004}, we can find that
\begin{align} \label{RNMV}
    ds^2 = -\left[\frac{1-\frac{M^2}{4a(t)^2r^2}+\frac{(Q_\text{e}^2 + Q_\text{m}^2)e^{-\gamma}}{4a(t)^2r^2}}{\left(1+\frac{M}{2a(t)r}\right)^2-\frac{(Q_\text{e}^2 + Q_\text{m}^2)e^{-\gamma}}{4a(t)^2r^2}}\right]^2 dt^2
    + a(t)^2 \left[\left(1+\frac{M}{2a(t)r}\right)^2-\frac{(Q_\text{e}^2 + Q_\text{m}^2)e^{-\gamma}}{4a(t)^2r^2}\right]^2 (dr^2 + r^2 d\Omega^2),
\end{align}
where $d\Omega^2 = \sin^2 \vartheta d\varphi^2 + d\vartheta^2$, and $a(t) = e^{H_\text{0}t}$ is the scale factor. In addition, $H_\text{0}$ is the present value of the Hubble constant. The relation between the cosmological constant $\Lambda$ and $H_\text{0}$ is
\begin{equation}
    \Lambda = 3 H^2_\text{0}.
\end{equation}
At present, $\Lambda = 1.1056$x$10^{-52}\text{ m}^{-2}$, which makes the distance of the cosmological horizon to be $d_\text{cosmo} = 9.51$x$10^{25}\text{ m}$. Note the time dependence of Eq. \eqref{RNMV}, but such dependence is negligible if one is near the black hole. Being near means comparing the distance of the black hole to the observer and the cosmological horizon. Indeed, even if we are so far to M87* or Sgr. A*, our distances from these black holes are so small compared to the scale of the cosmological horizon. Then by \cite{Tsupko2020}, $t_\text{0}-t<<H_\text{0}^{-1}$. Thus, as $a(t) \sim a(t_\text{0}) =$constant and using $x = a(t_\text{0}) r$, we can recast Eq. \eqref{RNMV} as 
\begin{align} \label{RNiso}
    ds^2 = -\left[\frac{1-\frac{M^2}{4x^2}+\frac{(Q_\text{e}^2 + Q_\text{m}^2)e^{-\gamma}}{4x^2}}{\left(1+\frac{M}{2x}\right)^2-\frac{(Q_\text{e}^2 + Q_\text{m}^2)e^{-\gamma}}{4x^2}}\right]^2 dt^2
    + \left[\left(1+\frac{M}{2x}\right)^2-\frac{(Q_\text{e}^2 + Q_\text{m}^2)e^{-\gamma}}{4x^2}    \right]^2 (dx^2 + x^2 d\Omega^2),
\end{align}
which is the RN black hole within isotropic coordinates. Introducing
\begin{equation} \label{e60}
    R = x\left(1+ \frac{2M}{x} - \frac{(Q_\text{e}^2 + Q_\text{m}^2)e^{-\gamma}}{x^2}\right)^2,
\end{equation}
where in the weak field limit ($M, Q_\text{e}, Q_\text{m} \sim 0$), $R \sim x$, we recover a form similar to the RN metric:
\begin{align} \label{e61}
    ds^2=-\left(1- \frac{2M}{R} + \frac{(Q_\text{e}^2 + Q_\text{m}^2)e^{-\gamma}}{R^2}\right)dt^2
    + \frac{dR^2}{\left(1- \frac{2M}{R} + \frac{(Q_\text{e}^2 + Q_\text{m}^2)e^{-\gamma}}{R^2}\right)} + R^2 d\Omega^2.
\end{align}

Let the present time be $t_\text{0}$ where the observer is at some radial position $r_\text{in}$, who observes the black hole shadow in the strong field regime. Here, the subscript "in" indicates the inner region. Then we should have $x_\text{in} = a(t_\text{0}) r_\text{in}$ and using these, Eq. \eqref{e60} can be recast as
\begin{equation} \label{e62}
    R_\text{in} = x_\text{in}\left(1+ \frac{2M}{x_\text{in}} - \frac{(Q_\text{e}^2 + Q_\text{m}^2)e^{-\gamma}}{x_\text{in}^2}\right)^2.
\end{equation}
Then using Eq. \eqref{e61}, an observer co-moving with the spacetime in Eq. \eqref{RNiso} will then observe the shadow radius $\mathcal{R}_\text{in}$ as
\begin{equation} \label{e63}
    \mathcal{R}_\text{in}=R_\text{in}\sin\alpha_\text{comov}=r_\text{ph}\left[\frac{2\left(1-\frac{2M}{R_\text{in}}+\frac{(Q_\text{e}^2 + Q_\text{m}^2)e^{-\gamma}}{R_\text{in}^{2}}\right)}{1-\frac{M}{r_\text{ph}}}\right]^{1/2},
\end{equation}
where $r_\text{ph}$ is still given by Eq. \eqref{erph}. We note that Eq. \eqref{e63} is the "inner" solution to the shadow radius where the effect of the cosmological expansion is considered negligible.

Next, we consider the solution to the outer region in the weak field regime of the black hole while considering a stronger effect of the cosmological expansion. It is easy to see how Eq. \eqref{RNMV} will reduce to an FRW metric in such a case:
\begin{equation} \label{FRW}
    ds^2 = -dt^2 + a(t)^2 (dr^2 + r^2 d\Omega^2).
\end{equation}
In this spacetime, it is well-known that the effective linear shadow radius $L_\text{sh}$, now applied to the shadow, is given in terms of the angular size of the black hole shadow $\Psi_\text{cosmo}$ as
\begin{equation} \label{e65}
    L_\text{sh} = \Psi_\text{cosmo} D_\text{A}(z),
\end{equation}
where
\begin{equation} \label{e66}
    D_\text{A}(z) = \frac{1}{1+z}\int_\text{0}^z \frac{dk}{H(k)},
\end{equation}
and
\begin{equation}
    H(k) = H_\text{0}[\Omega_\text{mat}(1+k)^3+\Omega_\text{rad}(1+k)^4+\Omega_\Lambda]^{1/2}.
\end{equation}
Here, $\Omega_\text{mat}, \Omega_\text{rad}, \Omega_\Lambda$ are present dimensionless density parameters for matter, radiation and dark energy, respectively. The angle $\Psi_\text{cosmo}$ is considered to be so tiny that the relation $\sin(\Psi_\text{cosmo}) \sim \Psi_\text{cosmo}$ remains valid. Due to the dependence of Eq. \eqref{e65} $z$, we can assume that $x_\text{out} \sim R_\text{out}$ due to the enormous distance from the black hole. It enables one to connect $z$ to $R_\text{out}$ via by \cite{Tsupko2020}
\begin{equation} \label{e68}
    R_\text{out} = \int_\text{0}^z \frac{dk}{H(k)}.
\end{equation}

Between these inner and outer regions, there is a location where the influence of the black hole and cosmological expansion begins to lose and gain, respectively. This overlap region, denoted by $R_\text{o}$, is still remote from the black hole. Thus, the approximation $z<<1$ is valid since the scale factor is $a(t_\text{0})$ at the location of such an observer at time $t_\text{0}$. Thus, we have $D_\text{A}(z) \sim R_\text{o}$ and
\begin{equation} \label{e69}
    L_\text{o} = \Psi_\text{o} R_\text{o},
\end{equation}
which is still equal to the weak field approximation of Eq. \eqref{e63}:
\begin{equation} \label{e70}
    \mathcal{R}_\text{o} = L_\text{o}=3\sqrt{3}M+\frac{\sqrt{3}((Q_\text{e}^2 + Q_\text{m}^2)e^{-\gamma})^{2}}{2}\left[\frac{1}{M}+\frac{1}{R_\text{o}}\right]+\mathcal{O}(R_\text{o}^{-2},R_\text{o}^{-3}).
\end{equation}
Then, we can match $L_\text{o}$ and $L_\text{sh}$ \cite{Tsupko2020}. Using $R_\text{out}$ instead of $R_\text{o}$, the effective shadow radius $\mathcal{R}_\text{cosmo}$ in the outer region of rapid expansion is obtained:
\begin{align}
    \mathcal{R}_\text{cosmo} = \Psi_\text{cosmo}R_\text{out} =\frac{R_\text{out}}{D_\text{A}(z)}\Bigg[3\sqrt{3}M
    +\frac{\sqrt{3}\left(Q_\text{e}^2 + Q_\text{m}^2\right)e^{-2\gamma}}{2}\left(\frac{1}{M}+\frac{1}{R_\text{out}}\right)+\mathcal{O}(R_\text{out}^{-2},R_\text{out}^{-3})\Bigg].
\end{align}
Finally, the perceived shadow radius $\mathcal{R}_\text{approx}$ by an observer co-moving with the cosmic expansion is approximated via composite solution \cite{Tsupko2020}:
\begin{equation} \label{e72}
    \mathcal{R}_\text{approx} = \mathcal{R}_\text{in} + \mathcal{R}_\text{cosmo} - \mathcal{R}_\text{o}.
\end{equation}
Note that in Ref. \cite{Tsupko2020}, the authors analyzed the situation using the angular radius, whereas, in this study, we slightly extended the formulation to investigate the shadow radius behaviour.

Eq. \eqref{e72} enables us to calculate the approximated shadow radius using different models for our Universe. In particular, let us take, for example, a Universe dominated by dark energy, where, after evaluating Eqs. \eqref{e66} and \eqref{e68}, we find
\begin{equation}
    D_\text{A}(z) = \frac{z}{(1+z)H_\text{0}}, \quad R_\text{out} = \frac{z}{H_\text{0}}.
\end{equation}
Now for brevity, let
\begin{align}
    W = r_\text{ph}\left[\frac{2\left(1-\frac{2M}{R_\text{out}}+\frac{\left(Q_\text{e}^2 + Q_\text{m}^2\right)e^{-\gamma}}{R_\text{out}^2}\right)}{1-\frac{M}{r_\text{ph}}} \right]^{1/2}, \qquad
    w = 3\sqrt{3}M+\frac{\sqrt{3}\left(Q_\text{e}^2 + Q_\text{m}^2\right)e^{-2\gamma}}{2}\left[\frac{1}{M}+\frac{1}{R_\text{out}}\right].
\end{align}
Then, Eq. \eqref{e72} implies that the approximate shadow radius seen by a co-moving observer is
\begin{equation} \label{e75}
    \mathcal{R}_\text{approx}^{\Lambda} = W + w H_\text{0} R_\text{out}.
\end{equation}
Consider a Universe dominated by matter where
\begin{equation}
    D_\text{A}(z) = \frac{2\left(\sqrt{z+1}-1\right)}{H_\text{0}(z+1)^{3/2}}, \quad R_\text{out} = \frac{2\left(\sqrt{z+1}-1\right)}{H_\text{0}\sqrt{z+1}},
\end{equation}
and we find
\begin{align} \label{e77}
    \mathcal{R}_\text{approx}^\text{mat} = W + w\left[\left(1-\frac{H_\text{0} R_\text{out}}{2}\right)^{-2}-1\right].
\end{align}
Lastly, for the radiation dominated Universe,
\begin{equation}
    D_\text{A}(z) = \frac{z}{(z+1)^2 H_\text{0}}, \quad R_\text{out} = \frac{z}{(z+1) H_\text{0}},
\end{equation}
and
\begin{align} \label{e79}
    \mathcal{R}_\text{approx}^\text{rad} = W - w \left[1 + (R_\text{out} H_\text{0} - 1)^{-1} \right].
\end{align}
The shadow radius behaviour seen by an observer in these models of the Universe can be visualized by plotting Eqs. \eqref{e75}, \eqref{e77}, \eqref{e79} numerically for an immediate comparison.
\begin{figure}
   \centering
    \includegraphics[width=0.48\textwidth]{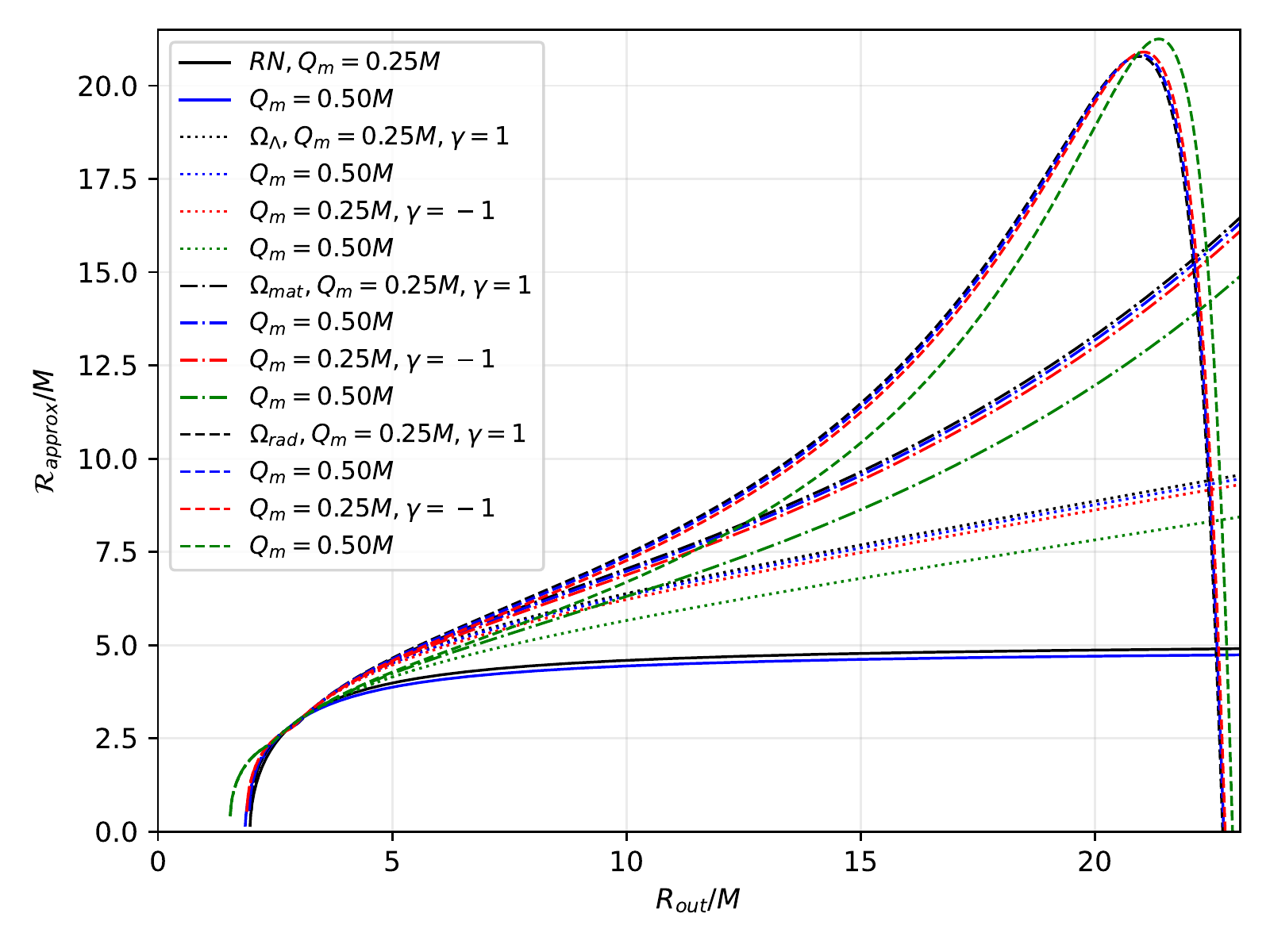}
-    \caption{Plot of the shadow radius for an observer co-moving with cosmological expansion in dark energy, matter, and radiation dominated Universes. Here, $Q_\text{m} = 0.10M$, and $H_\text{0} = 0.0408 \text{ M}^{-2}$.}
    \label{shacom}
\end{figure}
In Fig. \eqref{shacom}, we note that the Hubble constant is scaled to see the cosmological expansion's overall effect at short distances. However, it is understood that without this scaling and if one uses the experimental value of $H_\text{0}$, the effect can only be noticeable at distances near the cosmological horizon. Note that at low values of $R_\text{obs}/M$, we can see that the difference between these models is almost negligible. However, for large values of $R_\text{obs}/M$, we can observe the differences between these models of the Universe. Worth noting is the radiation-dominated Universe since there is some point where a peak is reached, and the shadow rapidly decreases in size for a very small change in $R_\text{obs}/M$. For dark energy and matter-dominated Universe, the co-moving observer sees no limit on large the shadow radius will increase. As a final remark, it can also be observed in the plot that the effect of the screening parameter $\gamma$ is affected by the co-moving state of motion and the type of Universe the observer lives in.

\section{Deflection Angle Using Gauss-Bonnet Theorem In Weak Field Limits} \label{sec3}
In this section, we aim to investigate the weak deflection angle of both massive and null particles by exploiting the Gauss-Bonnet theorem (GBT), which is originally stated as \cite{o1992riemannian,Gibbons:2008rj},
\begin{equation} \label{eGBT}
    \iint_DKdS+\sum\limits_{a=1}^N \int_{\partial D_{a}} \kappa_{\text{g}} d\ell+ \sum\limits_{a=1}^N \theta_{a} = 2\pi\chi(D).
\end{equation}
In this equation, the Gaussian curvature $K$ that describes the domain $D$ can be oriented in a $2D$ curved surface $S$ with infinitesimal area element $dS$. The boundaries of $D$ are given by $\partial D_{\text{a}}$ (a=$1,2,..,N$), and the geodesic curvature $\kappa_{\text{g}}$ is integrated over the path $d\ell$ in a positive sense. Also, the jump angle is denoted by $\theta_\text{a}$, wherein $\chi(D)$ is the Euler characteristic equal to $1$ due to the non-singularity of $D$.

In Ref. \cite{Li:2020wvn}, the GBT is applied to accommodate black hole metrics with non-asymptotic flatness. Although our metric is asymptotically flat, we can still use this formalism as a special case. With the photonsphere radius $r_{co}$ as being part of the quadrilateral, which serves as the domains of integration, the weak deflection angle can be derived via
\begin{equation} \label{eIshi}
    \hat{\alpha} = \iint_{_{r_\text{co}}^{R }\square _{r_\text{co}}^{S}}KdS + \phi_{\text{RS}},
\end{equation}
where S and R are the radial positions of the source and receiver, respectively. Furthermore, the infinitesimal curve surface $dS$ is given by
\begin{equation}
    dS = \sqrt{g}drd\phi,
\end{equation}
and $\phi_\text{RS}$ is the coordinate position angle between the source and the receiver defined as $\phi_\text{RS} = \phi_\text{R}-\phi_\text{S}$, and $g$ is the determinant of the Jacobi metric. line element of a static, spherically symmetric (SSS) spacetime
\begin{align}
    ds^2=g_{\mu \nu}dx^{\mu}dx^{\nu}=-A(r)dt^2+B(r)dr^2+C(r)d\Omega^2,
\end{align}
where $d\Omega^2=d\theta^2+\sin^2\theta d \phi^2$, the Jacobi metric is defined as 
\begin{align} \label{eJac}
    dl^2=g_{ij}dx^{i}dx^{j}
    =(E^2-\mu^2A(r))\left(\frac{B(r)}{A(r)}dr^2+\frac{C(r)}{A(r)}d\Omega^2\right).
\end{align}
Here, $E$ is the energy of the massive particle defined by
\begin{equation} \label{en}
    E = \frac{\mu}{\sqrt{1-v^2}},
\end{equation}
where $v$ is the particle's velocity. Due to spherical symmetry, we can analyze along the equatorial plane without losing generality. The determinant of the Jacobi metric can then be sought off as
\begin{equation}
    g=\frac{B(r)C(r)}{A(r)^2}(E^2-\mu^2 A(r))^2.
\end{equation}
We need the orbit equation to determine the expressions for $\phi_\text{RS}$. For time-like particles, this can be derived through $g_{\mu \nu}dx^{\mu}dx^{\nu} = -1$ which results to
\begin{align}
    F(u) \equiv \left(\frac{du}{d\phi}\right)^2 
    = \frac{C(u)^2u^4}{A(u)B(u)}\Bigg[\left(\frac{E}{J}\right)^2-A(u)\left(\frac{1}{J^2}+\frac{1}{C(u)}\right)\Bigg],
\end{align}
where we have used the substitution $r = 1/u$ and the angular momentum of the massive particle
\begin{equation}
    J = \frac{\mu v b}{\sqrt{1-v^2}},
\end{equation}
and $b$ is the impact parameter. With the metric coefficients, we find
\begin{align}
    F(u) = \frac{E^2-1}{J^2}-u^2-u^2\left(\frac{1}{J^2}+u^2\right)\left(Q_\text{e}^2 + Q_\text{m}^2\right)e^{-\gamma}
    + \left(\frac{1}{J^2}+u^2\right)2Mu.
\end{align}
Solving the above iteratively, we find
\begin{equation} \label{orb}
    u(\phi) = \frac{\sin(\phi)}{b}+\frac{1+v^2\cos^2(\phi)}{b^2v^2}M - \frac{\left(Q_\text{e}^2 + Q_\text{m}^2\right)e^{-\gamma}}{2v^2 b^3}.
\end{equation}

The Gaussian curvature $K$, in terms of affine connections and determinant $g$, is defined as
\begin{align}
    K=\frac{1}{\sqrt{g}}\left[\frac{\partial}{\partial\phi}\left(\frac{\sqrt{g}}{g_{rr}}\Gamma_{rr}^{\phi}\right)-\frac{\partial}{\partial r}\left(\frac{\sqrt{g}}{g_{rr}}\Gamma_{r\phi}^{\phi}\right)\right] 
    =-\frac{1}{\sqrt{g}}\left[\frac{\partial}{\partial r}\left(\frac{\sqrt{g}}{g_{rr}}\Gamma_{r\phi}^{\phi}\right)\right]
\end{align}
since $\Gamma_{rr}^{\phi} = 0$ for Eq. \eqref{eJac}. Then with the analytical solution to $r_\text{co}$,
\begin{equation}
    \left[\int K\sqrt{g}dr\right]\bigg|_{r=r_\text{co}} = 0,
\end{equation}
thus,
\begin{equation} \label{gct}
    \int_{r_\text{co}}^{r(\phi)} K\sqrt{g}dr = -\frac{A(r)\left(E^{2}-A(r)\right)C'-E^{2}C(r)A(r)'}{2A(r)\left(E^{2}-A(r)\right)\sqrt{B(r)C(r)}}\bigg|_{r = r(\phi)}.
\end{equation}
The prime denotes differentiation with respect to $r$. The weak deflection angle is then \cite{Li:2020wvn},
\begin{align} \label{eqwda}
    \hat{\alpha} = \int^{\phi_\text{R}}_{\phi_\text{S}} \left[-\frac{A(r)\left(E^{2}-A(r)\right)C'-E^{2}C(r)A(r)'}{2A(r)\left(E^{2}-A(r)\right)\sqrt{B(r)C(r)}}\bigg|_{r = r(\phi)}\right] d\phi + \phi_\text{RS}.
\end{align}
Using Eq. \eqref{orb} in Eq. \eqref{gct}, we find
\begin{align} \label{gct2}
    &\left[\int K\sqrt{g}dr\right]\bigg|_{r=r_\phi} = -\frac{\left(2E^{2}-1\right)M(\cos(\phi_\text{R})-\cos(\phi_\text{S}))}{\left(E^{2}-1\right)b} \nonumber\\
    &-\frac{\left(3E^{2}-1\right)(\left(Q_\text{e}^2 + Q_\text{m}^2\right)e^{-\gamma}\left[\phi_{RS}-\frac{(\sin(2\phi_\text{R})-\sin(2\phi_\text{S})}{2}\right]}{4\left(E^{2}-1\right)b^{2}}
     -\phi_{RS}
    + \mathcal{O}[M\left(Q_\text{e}^2 + Q_\text{m}^2\right)e^{-\gamma}].
\end{align}
Next, we find the expression for $\phi$. To do this, we use Eq. \eqref{orb} and solve for $\phi$. For the source and receiver, we find
\begin{align} \label{s}
    \phi_\text{S} =\arcsin(bu)+\frac{M\left[v^{2}\left(b^{2}u^{2}-1\right]-1\right)}{bv^{2}\sqrt{1-b^{2}u^{2}}} 
    +\frac{\left(Q_\text{e}^2 + Q_\text{m}^2\right)e^{-\gamma}}{2b^{2}v^{2}\sqrt{1-b^{2}u^{2}}} 
    + \mathcal{O}[M\left(Q_\text{e}^2 + Q_\text{m}^2\right)e^{-\gamma}],
\end{align}
\begin{align} \label{r}
    \phi_\text{R} =\pi -\arcsin(bu)-\frac{M\left[v^{2}\left(b^{2}u^{2}-1\right]-1\right)}{bv^{2}\sqrt{1-b^{2}u^{2}}} 
    -\frac{\left(Q_\text{e}^2 + Q_\text{m}^2\right)e^{-\gamma}}{2b^{2}v^{2}\sqrt{1-b^{2}u^{2}}} 
    + \mathcal{O}[M\left(Q_\text{e}^2 + Q_\text{m}^2\right)e^{-\gamma}]
\end{align}
respectively. Careful observation of these equations will allow us to write $\phi_\text{RS} = \pi - 2\phi_\text{S}$. Now, we take note of the following relations:
\begin{align}
    \cos(\pi-\phi_\text{S})=-\cos(\phi_\text{S}), \qquad
    \cot(\pi-\phi_\text{S})=-\cot(\phi_\text{S}), \qquad
    \sin(\pi-\phi_\text{S})=\sin(\phi_\text{S}).
\end{align}
The last property cancels the sine terms in Eq. \eqref{gct2}. We find $\cos(\phi_\text{S})$ as
\begin{align} \label{cs}
    \cos(\phi_\text{S}) = \sqrt{1-b^{2}u^{2}}-\frac{Mu\left[v^{2}\left(b^{2}u^{2}-1\right)-1\right]}{v^{2}\sqrt{\left(1-b^{2}u^{2}\right)}}
    -\frac{\left(Q_\text{e}^2 + Q_\text{m}^2\right)e^{-\gamma}u}{\sqrt{2}\sqrt{bv^{2}\left(1-b^{2}u^{2}\right)}}
    + \mathcal{O}[M\left(Q_\text{e}^2 + Q_\text{m}^2\right)e^{-\gamma}],
\end{align}
and $\cot(\phi_\text{S})$ as
\begin{align} \label{cr}
    \cot(\phi_\text{S}) = \frac{\sqrt{1-b^{2}u^{2}}}{bu}+\frac{M\left[v^{2}(-b^{2}u^{2}+1)+1\right]}{b^{3}u^{2}v^{2}\sqrt{1-b^{2}u^{2}}}
    -\frac{\left(Q_\text{e}^2 + Q_\text{m}^2\right)e^{-\gamma}}{2b^{4}u^{2}v^{2}\sqrt{1-b^{2}u^{2}}}
    + \mathcal{O}[M\left(Q_\text{e}^2 + Q_\text{m}^2\right)e^{-\gamma}].
\end{align}
Using Eq. \eqref{en} and by plugging Eqs. \eqref{s}-\eqref{cr} in Eq. \eqref{eqwda}, we finally obtain
\begin{align} \label{ewda}
    \hat{\alpha} &\sim \frac{M\left(v^{2}+1\right)}{bv^{2}}\left(\sqrt{1-b^{2}u_\text{R}^{2}}+\sqrt{1-b^{2}u_\text{S}^{2}}\right)
    -\frac{\left(Q_\text{e}^2 + Q_\text{m}^2\right)e^{-\gamma}\left(v^{2}+2\right)}{4b^{2}v^{2}}\left[\pi-(\arcsin(bu_\text{R})+\arcsin(bu_\text{S}))\right] \nonumber \\
    &+\mathcal{O}[M\left(Q_\text{e}^2 + Q_\text{m}^2\right)e^{-\gamma}],
\end{align}
which also involves the finite distance $u_\text{S}$ and $u_\text{R}$. The above expression can still be further approximated as $b^2u^2 \sim 0$:
\begin{align} \label{ewda2}
    \hat{\alpha} \sim \frac{2M\left(v^{2}+1\right)}{bv^{2}}-\frac{\left(Q_\text{e}^2 + Q_\text{m}^2\right)e^{-\gamma} \pi\left(v^{2}+2\right)}{4b^{2}v^{2}}
    +\mathcal{O}[M\left(Q_\text{e}^2 + Q_\text{m}^2\right)e^{-\gamma}].
\end{align}
For the case of photons where $v=1$, we find
\begin{align} \label{ewda3}
    \hat{\alpha} \sim \frac{4M}{b}-\frac{3\pi\left(Q_\text{e}^2 + Q_\text{m}^2\right)e^{-\gamma} }{4b^{2}}
    +\mathcal{O}[M\left(Q_\text{e}^2 + Q_\text{m}^2\right)e^{-\gamma}].
\end{align}
The above agrees with the result in Ref. \cite{Ishihara2016}.
\begin{figure*}
    \centering
    \includegraphics[width=0.48\textwidth]{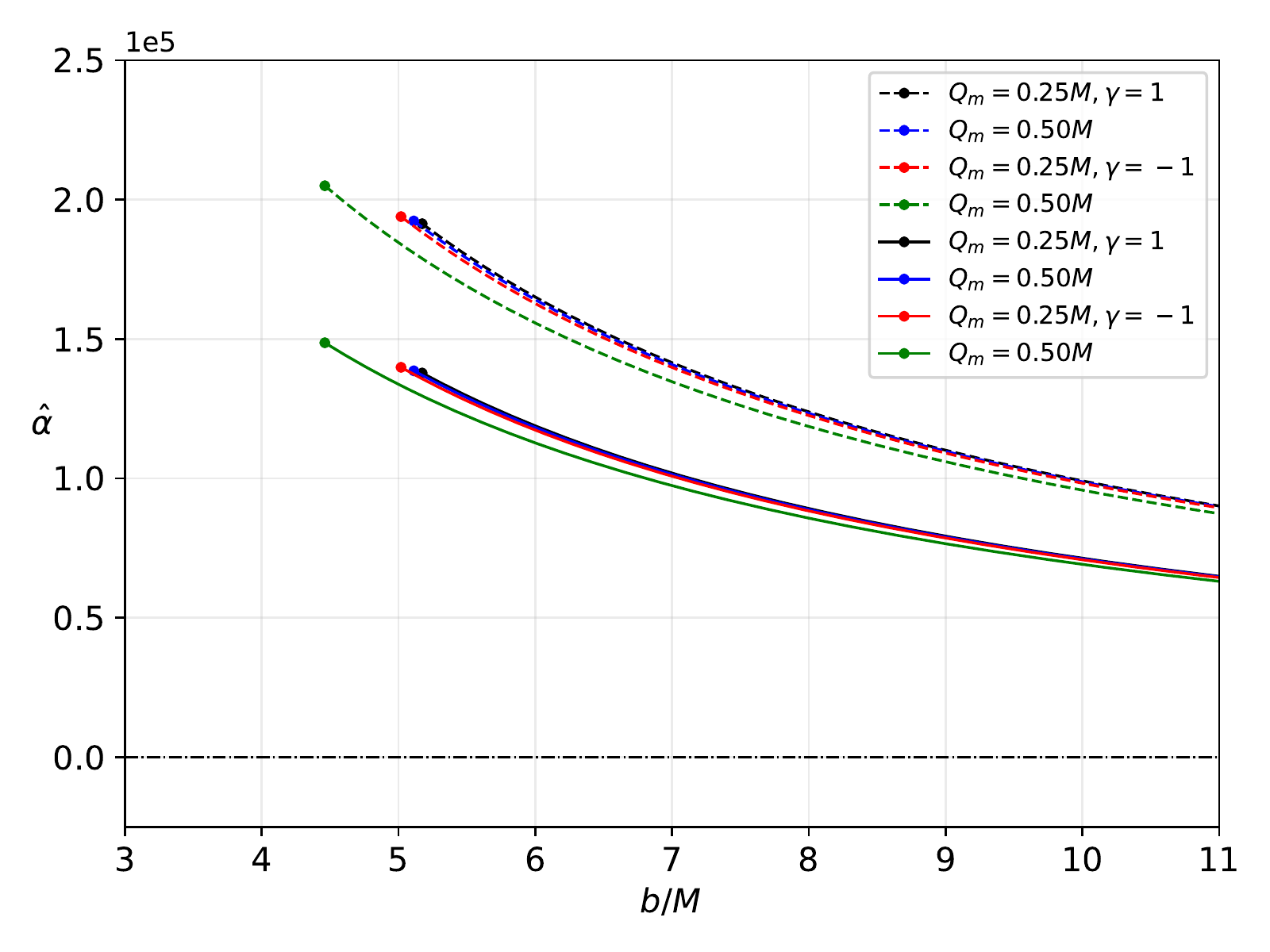}
    \includegraphics[width=0.48\textwidth]{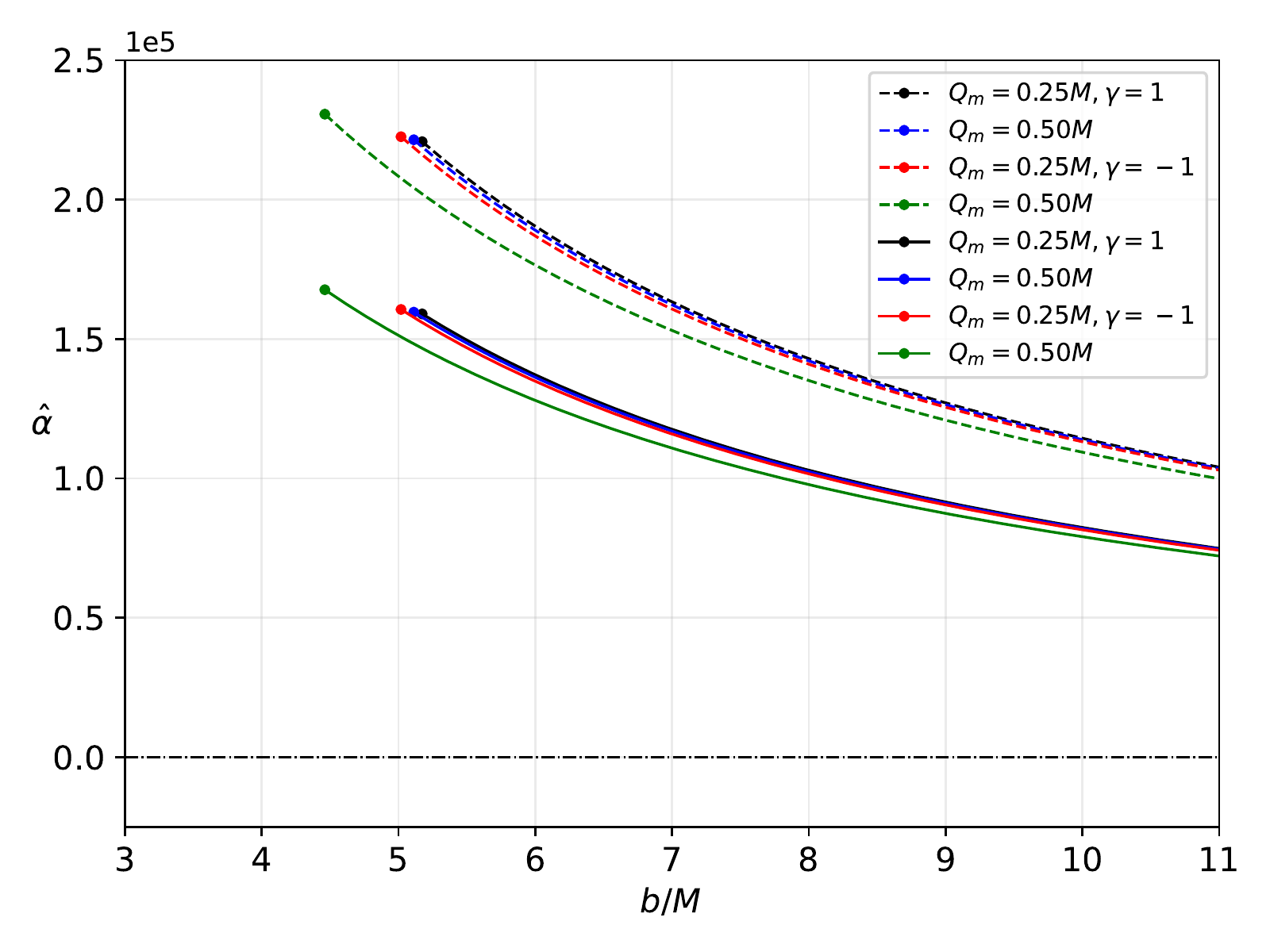}
    \caption{Left: Weak deflection angle with finite distance given by Eq. \eqref{ewda}. Here, $u_\text{S} = u_\text{R} = 0.5b^{-1}$. The dashed lines represent the massive particles, and the solid lines are for photons. Right: Weak deflection angle as given by Eq. \eqref{ewda2}, where $u_\text{S} = u_\text{R} \sim 10^{-10}b^{-1}$. In these plots, the dot represents the critical impact parameters in every case.}
    \label{wda}
\end{figure*}
How will the screening parameter $\gamma$ affect the weak deflection angle, in combination with the finite distance effects? These can be seen in Fig. \ref{wda}, where we also compared the $\hat{\alpha}$ caused by both massive and null particles. Our results indicate that $\hat{\alpha}$ is increased as both the source and the receiver are approximated at $r\to \infty$. Interestingly, massive particles also give a larger deflection angle than photons. When it comes to the effect of $\gamma$, it does not discriminate whether it is massive or null particles. That is, $\gamma>0$ has a larger value for $\hat{\alpha}$ as compared to $\gamma<0$. We also remark that $\gamma<0$ gives a much lower bound to the critical impact parameter. It means that near (but greater) than the critical impact parameter, $\hat{\alpha}$ has more range to exist.

A useful astronomical application of the weak deflection angle involves the Einstein ring. Let $D_\text{SL}$ and $D_\text{LR}$ be the position of the source and receiver from the black hole (lensing object $L$), respectively. The thin lens approximation implies that $D_\text{RS}=D_\text{SL}+D_\text{LR}$, and the position of the weak field images is given by
\begin{equation}
    D_\text{RS}\tan\beta=\frac{D_\text{LR}\sin\theta-D_\text{SL}\sin(\hat{\alpha}-\theta)}{\cos(\hat{\alpha}-\theta)}.
\end{equation}
An Einstein ring will be formed when $\beta=0$, and the above equation simplifies to
\begin{equation}
    \theta_\text{E}\sim\frac{D_\text{SL}}{D_\text{RS}}\hat{\alpha}.
\end{equation}
Finally, using the relation $b=D_\text{LR}\sin\theta \sim D_\text{LR}\theta$ we can obtain
\begin{align} \label{ering}
    \theta_\text{E} = \frac{-9\pi\epsilon D_\text{SL}D_\text{LR}}{4D_\text{LR}\left(\Lambda+6\right)\left(D_\text{RS}\right)}
    +\frac{\sqrt{3}D_\text{SL}D_\text{LR}\left\{ 128\left(\Lambda+6\right)\left(D_\text{RS}\right)M+27\pi^{2}D_\text{SL}D_\text{LR}\epsilon^{2 }\right\} ^{1/2}}{4D_\text{LR}\left(\Lambda+6\right)\left(D_\text{RS}\right)}
\end{align}
where the parameter $\epsilon=\left(Q_\text{e}^2 + Q_\text{m}^2\right)e^{-\gamma}/b^2$.
\begin{figure}
    \centering
    \includegraphics[width=0.48\textwidth]{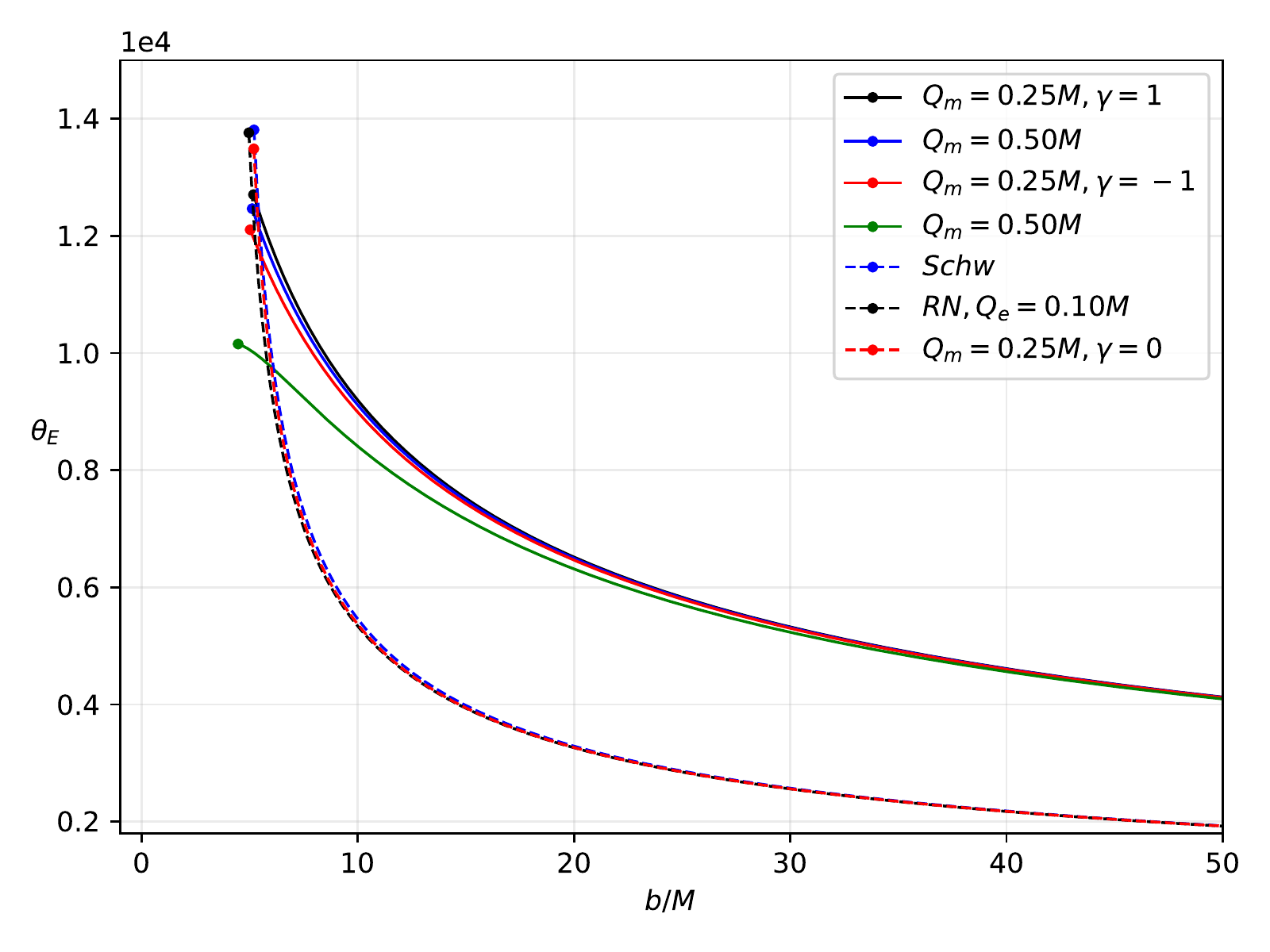}
    \caption{Plot (theoretical) of the Einstein ring formation in $\mu$as. Here, we set $Q_\text{e} = 0.10M$, and $D_\text{S} = D_\text{R} = 10^{10}b$. The dots represent the critical impact parameter.}
    \label{einsring}
\end{figure}
We plot Eq. \eqref{ering} as shown in Fig. \ref{einsring}. In this plot, we compared four cases, and the solid lines represent the effect of the screening parameter on the Einstein ring. As $\gamma$ decreases, it can be gleaned that the Einstein rings angular radius decreases for a given impact parameter. Furthermore, how fast these angular radii change with $b/M$ is less than cases with no screening parameter (dashed line). As $b/M \to \infty$, the Einstein rings approach the Schwarzschild case. Note that with these parameters, the angular radius of the formed Einstein ring can be experimentally detected since the lowest value we can get in the plot is $\sim 10000\mu$as. Furthermore, if the deviation from $Q_\text{m} = 0.25M$ and $Q_\text{m} = 0.50M$ at $\gamma = -1$ is examined, one needs a device capable of detailing a difference of $570\mu$as at $b/M = 10$. We see that as $b/M \to \infty$, the effect of the dyonic Modmax parameters lessens.

\section{Spherically infalling accretion} \label{sec4}
In this section, we study the spherically free-falling accretion model on the BH from infinity using the method defined in \cite{Jaroszynski:1997bw,Bambi:2012tg}. This method gives us realistic visualization of the shadow cast with the accretion disk. In reality, the actual image of the BH can not be seen as an apparent boundary in the universe. Moreover, it is not realistic to use a static accretion disk model because there is a moving accretion disk around a black hole, and also it provides a synchrotron
emission from the accretion. To do so, we first study the specific intensity observed at the observed photon frequency $\nu_\text{obs}$ solving this integral along the light ray:
        \begin{equation}
            I(\nu_\text{obs},b_\gamma) = \int_\gamma g^3 j(\nu_e) dl_\text{prop}.
            \label{eq:bambiI}
        \end{equation}
        
It is noted that $b_{\gamma}$  is the impact parameter, $j(\nu_e)$ is the emissivity/volume, $dl_\text{prop}$ is the infinitesimal proper length and $\nu_e$  stands for the photon frequency of the emitter. Here we define the redshift factor for the infalling accretion as follows:
        \begin{equation}
            g = \frac{k_\mu u^\mu_o}{k_\mu u^\mu_e},
        \end{equation}
where the 4-velocity of the photon is $k^\mu=\dot{x}_\mu$ and 4-velocity of the distant observer is $u^\mu_o=(1,0,0,0)$. Moreover, the $u^\mu_e$ stands for the 4-velocity of the infalling accretion
\begin{equation}
u_{\mathrm{e}}^{t}=\frac{1}{A(r)}, \quad u_{\mathrm{e}}^{r}=-\sqrt{\frac{1-A(r)}{A(r) B(r)}}, \quad u_{\mathrm{e}}^{\theta}=u_{\mathrm{e}}^{\phi}=0.
\end{equation}

Using the relation of  $k_{\alpha} k^{\alpha}=0$, one can derive $k_{r}$ and $k_{t}$ which is a constant of motion for the photons:
\begin{equation}
k_{r}=\pm k_{t} \sqrt{B(r)\left(\frac{1}{A(r)}-\frac{b^{2}}{r^{2}}\right)}.
\end{equation}
Note that sign of $\pm$ shows that the photon
gets close to (away from) the black hole. Then the redshift factor $g$ and proper distance $dl_\gamma$ can be written as follows
   \begin{equation}
   g = \Big( u_e^t + \frac{k_r}{k_t}u_e^r \Big)^{-1},
  \end{equation}
  and
 \begin{equation}
  dl_\gamma = k_\mu u^\mu_e d\lambda = \frac{k^t}{g |k_r|}dr.
\end{equation}
We then consider only the monochromatic emission for the specific emissivity with rest-frame frequency $\nu_*$ as follows:
        \begin{equation}
            j(\nu_e) \propto \frac{\delta(\nu_e - \nu_*)}{r^2}.
        \end{equation}
        
Afterwards, the intensity equation given in \eqref{eq:bambiI} becomes
        
        \begin{equation}
            F(b_\gamma) \propto \int_\gamma \frac{g^3}{r^2} \frac{k_e^t}{k_e^r} dr.
        \end{equation}
        
\begin{figure*}
    \centering
    \begin{tikzpicture}[every node/.style={anchor=center,inner sep=0pt},x=1mm, y=0mm,]   
    \node[opacity=0.80] (fig1) at (0,0)
    {\includegraphics[width=0.48\textwidth]{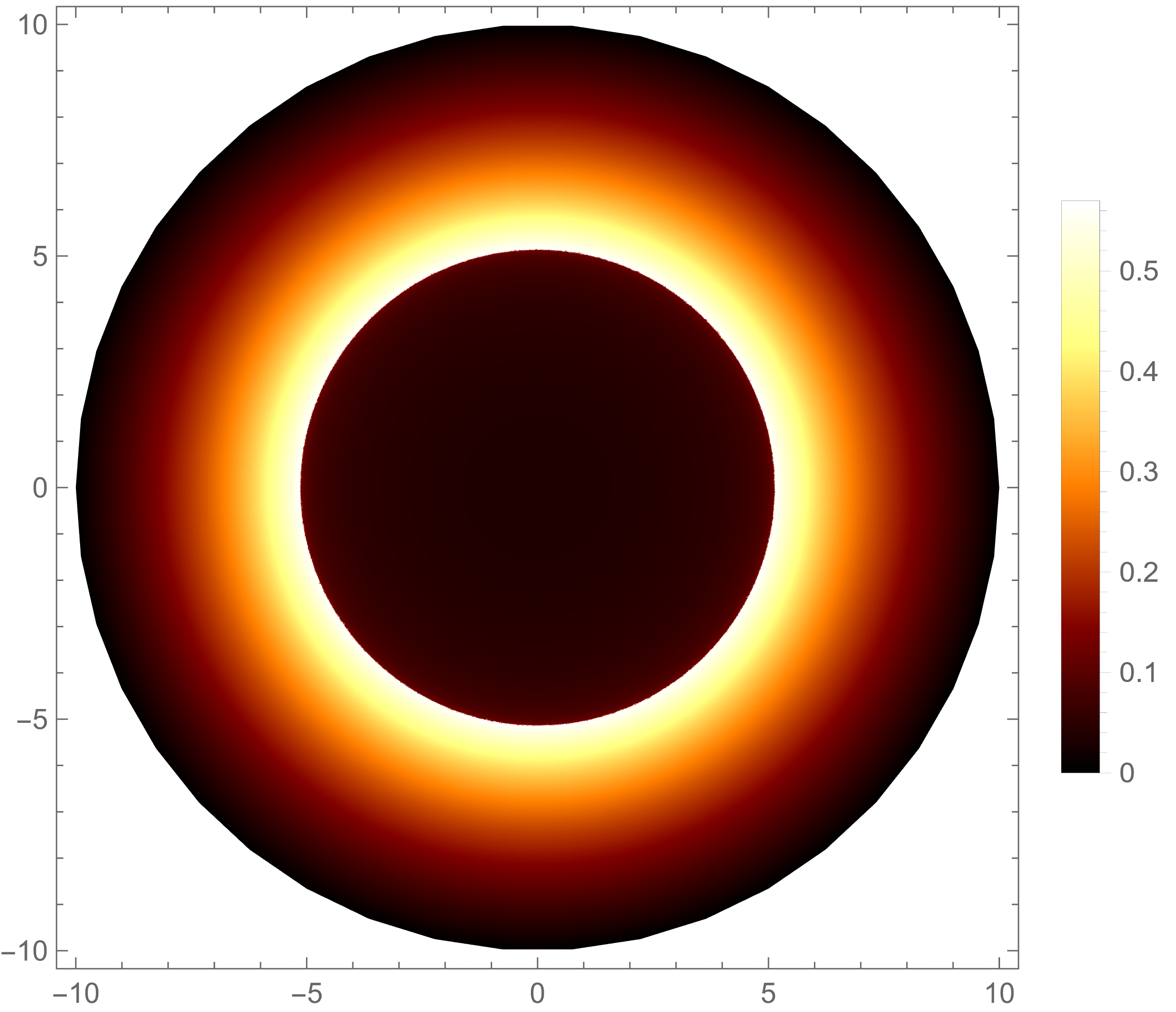}};
    \node (fig2) at (-2.1,0)
    {\includegraphics[width=7.5cm,height=12cm,keepaspectratio]{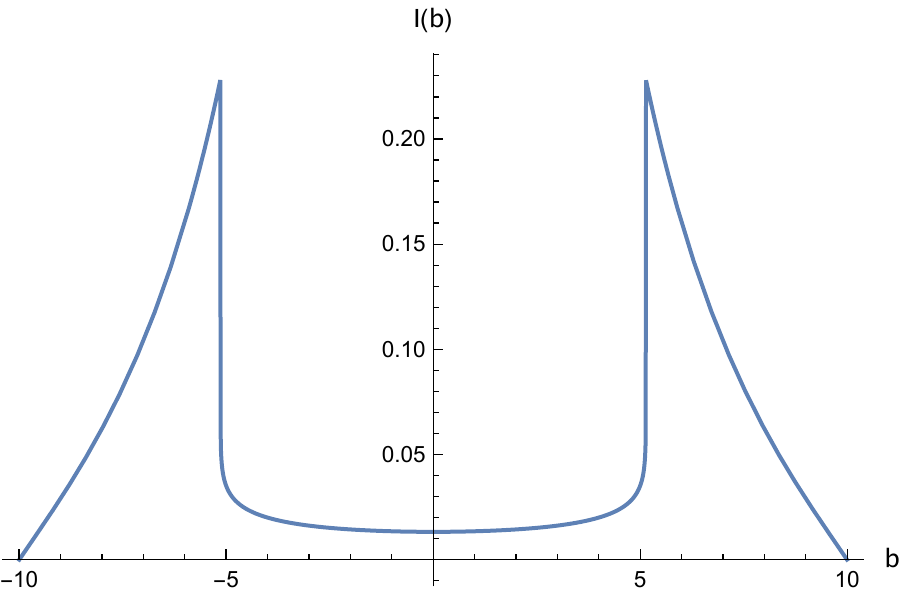}};  
    \end{tikzpicture}
    \caption{The specific intensity $I_\text{obs}$ seen by a distant observer for an infalling accretion at fixed $M=1$, $Q_\text{e}=Q_\text{m}=0.2$, and $\gamma=0.2m$.}
    \label{fig:thinacc1}	
\end{figure*}

\begin{figure*}
    \centering
    \begin{tikzpicture}[every node/.style={anchor=center,inner sep=0pt},x=1mm, y=0mm,]   
    \node[opacity=0.80] (fig1) at (0,0)
    {\includegraphics[width=0.48\textwidth]{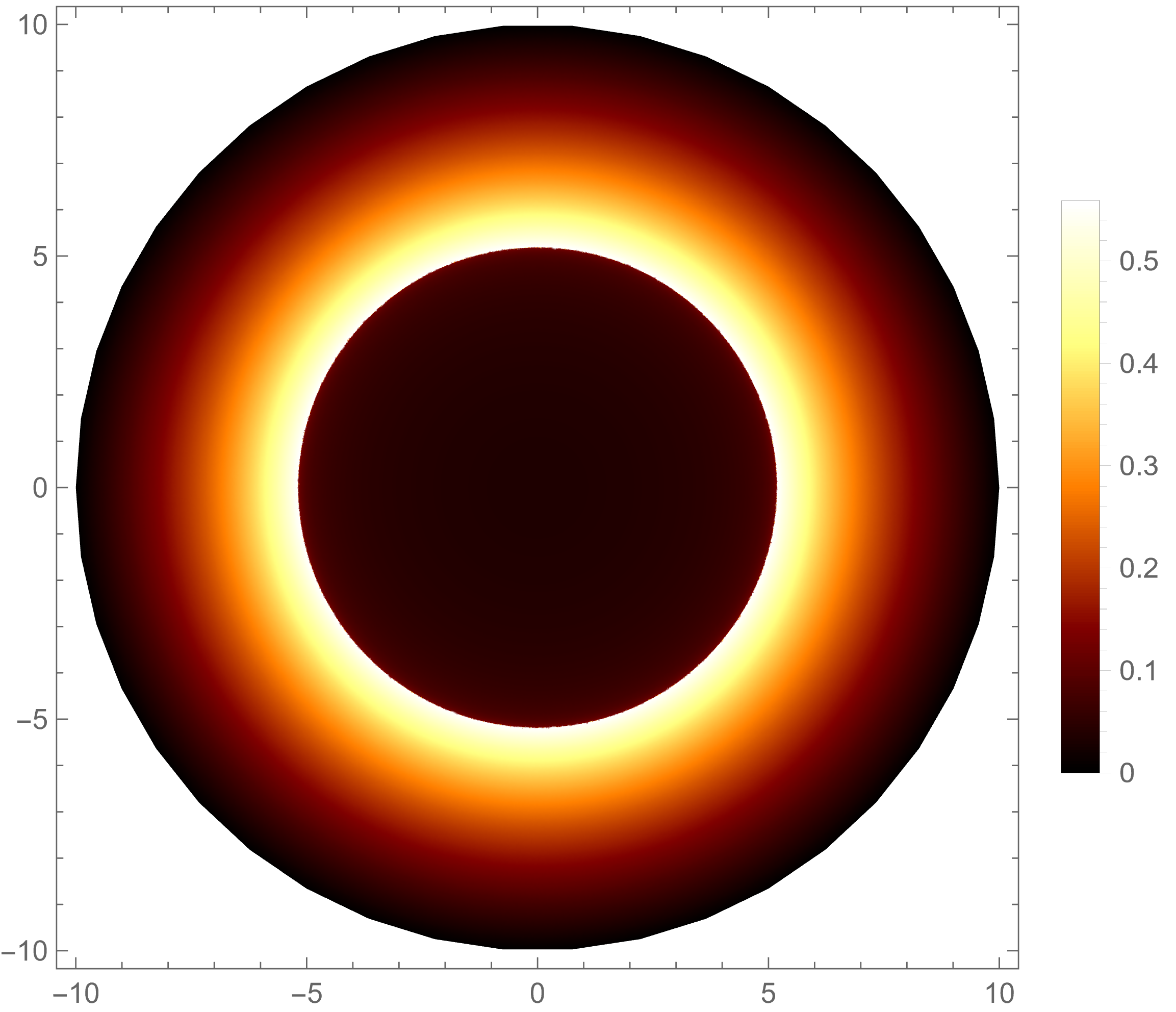}};
    \node (fig2) at (-2.1,0)
    {\includegraphics[width=7.5cm,height=12cm,keepaspectratio]{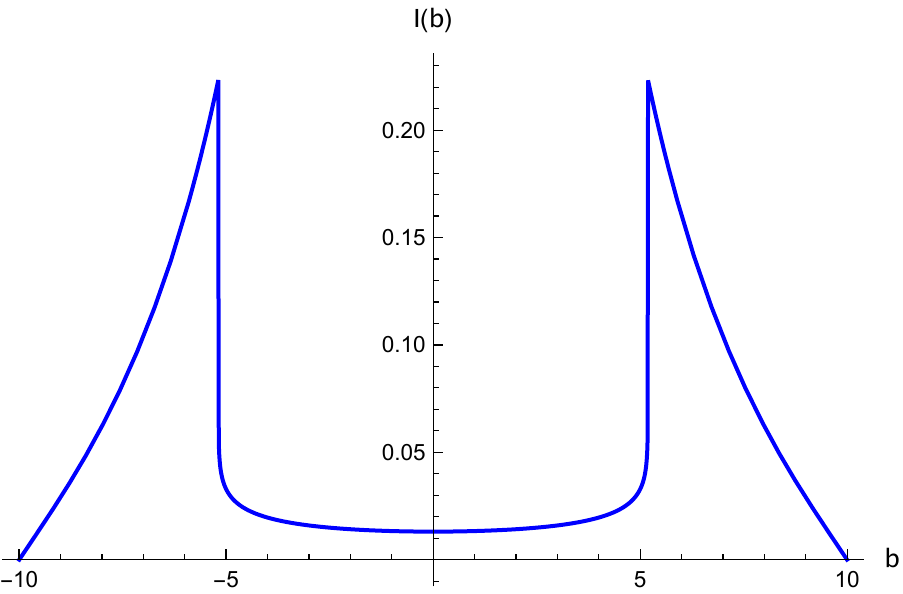}};  
    \end{tikzpicture}
    \caption{The specific intensity $I_\text{obs}$ seen by a distant observer for an infalling accretion at fixed $M=1$, $Q_\text{e}=Q_\text{m}=0.2$, and $\gamma=5m$.}
    \label{fig:thinacc2}	
\end{figure*}

\begin{figure}
    \centering
    \includegraphics[width=0.48\textwidth]{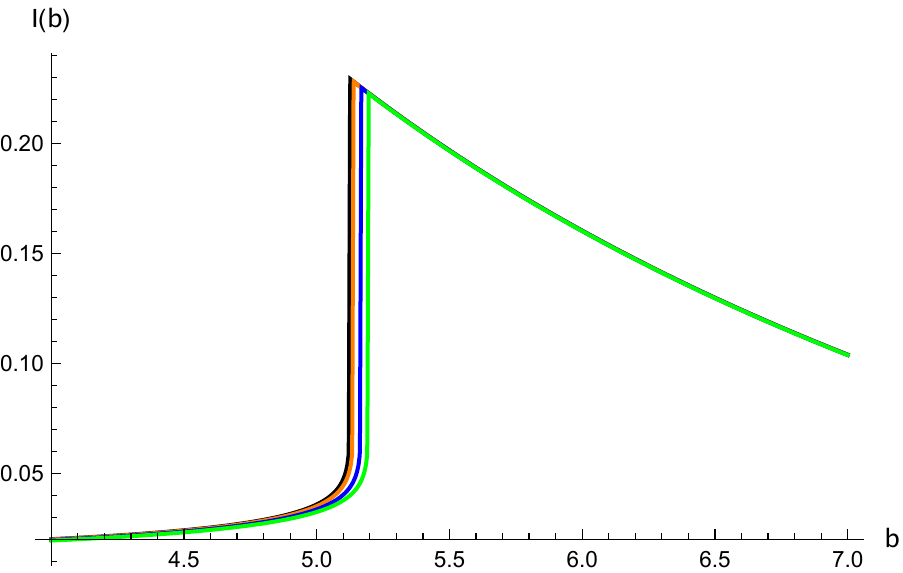}\hspace{1cm}
    \caption{The specific intensity $I_\text{obs}$ seen by a distant observer for an infalling accretion at fixed $M=1$, $Q_\text{e}=Q_\text{m}=0.2$, $\gamma=0m $(black),
    $\gamma=0.2m $(orange), $\gamma=0.9m $(blue) and $\gamma=5m $(green).}
    \label{fig:thinacc3}	
\end{figure}

We investigate the shadow cast with the thin-accretion disk of the black hole in ModMax. First of all, we solve the above equation numerically using the \textit{Mathematica} notebook package \cite{Okyay:2021nnh}, (also used in \cite{Chakhchi:2022fls,Kuang:2022xjp,Uniyal:2022vdu,Pantig:2022ely}) and this integration of the flux show the effects of the parameters of the ModMax $\gamma$ on the specific intensity seen by a distant observer for an infalling accretion in Figs. (\ref{fig:thinacc1}, \ref{fig:thinacc2} and \ref{fig:thinacc3}). These plots in Figs. (\ref{fig:thinacc1}, \ref{fig:thinacc2} and \ref{fig:thinacc3}) show the specific intensities for various values of the parameter $\gamma$ versus $b$ observed by the distant observer. 

We observe that increasing the value of $b$, increases the intensity first. Afterwards, intensity reaches the peak value sharply where the photons are captured by a black hole quickly ( at the photon sphere). It is seen that after the peak value, intensity slowly decreases. Moreover, we show the shadow cast of the black hole in the 2-dimensional image with a photon sphere by a distant observer in (X, Y) plane where the dark centre of it the event horizon is located, and it is circled by a bright ring with a strongest luminosity (photon sphere). It can be seen that brightness decreases gradually after the maximum region. Hence, we show the effect of the screening parameter $\gamma$ on the black hole luminosity of the shadow cast where the intensity decreases with increasing the value of the screening parameter $\gamma$ as seen in Fig. \ref{fig:thinacc3}.

\section{Quasinormal modes of Dyonic ModMax Black Holes} \label{sec5}
Since the first detection of gravitational waves (GWs) from the coalescence of two stellar-mass black holes in 2015 by the LIGO/VIRGO collaborations \cite{LIGOScientific:2016aoc}, gravitational wave physics has begun. Then data from gravitational waves are analyzed to test alternative theories of gravity and different models of compact objects. Perturbative analysis of black hole spacetimes dominated by `quasinormal ringing' is used to do this. Quasinormal modes (QNMs) are oscillations with complex frequencies with energy dissipation. The complex frequencies of QNMs have the characteristic properties of the BHs, such as mass, charge, and angular momentum, independent of the initial perturbations. 

In this section, we use the method of WKB approximation to study quasinormal modes (QNMs) of the Dyonic ModMax BHs. This method is known as an effective way to derive QNMs, which is firstly used in \cite{Schutz:1985km} by Schutz and Will, then Iyer and Will extended to the third order of WKB approximation in \cite{Iyer:1986np}. Recently, WKB approximation has been extended to  
sixth order and higher-order cases in \cite{Konoplya:2003ii,Konoplya:2019hlu}. To study QNMs of the Dyonic ModMax BHs, first, we use a massless scalar field
perturbation in the background of the black hole \ref{spherical-le}. The Klein-Gordon equation, which is used for the scalar field, can be written as follows:
\begin{equation}
\frac{1}{\sqrt{-g}}\partial_\mu\left( {\sqrt{-g}}g^{{\mu}{\nu}}\partial_\nu{\Phi} \right)=0,
\label{KG-wave-eq}
\end{equation}
where  $g$ is the determinant of the metric $g_{\mu\nu}$. Then to solve the above equation, we first use the separation of variables:
\begin{equation}
	\Phi(t,r, \theta, \phi) = Y_l(\theta, \phi)\Psi(t,r)/r.
	\label{seperate-variable}
\end{equation}

It is noted that $Y_l(\theta, \phi)$ are spherical harmonics with the multipole number $l=0,1,2,\dots$. We then obtain the Schrodinger-like wave equation (Regge-Wheeler-Zerilli  equations) \cite{Konoplya:2011qq,Kokkotas:1999bd}:
\beeq
\frac{\partial^2\Psi}{\partial t^2}+\left(-\frac{\partial^2}{\partial r_*^2}+V(r)\right)\Psi=0.
\label{WE-time}
\eneq  

where the following relatıon defines the tortoise coordinate $r_*$:
\beeq
dr_*=\frac{dr}{\sqrt{A(r)B(r)}},
\label{tortoise}
\eneq
with the effective potential for the scalar field
\beeq
V(r)=A(r) \frac{l(l+1)}{r^2}+ \frac{1}{2r} \frac{d}{dr} \left[ A(r)B(r) \right].
\label{scalarV}
\eneq

where the line element is written as:
\begin{equation}
ds^2=-A(r)dt^2+B(r)^{-1}dr^2+r^2d\Omega^2,
\label{spherical-le}
\end{equation} 

Note that $A(r)=B(r)=f(r)$ in Eq. \ref{eq:fdyon}.

One can think of the effective potentials as the potential barriers which provide decaying on the event horizon and at
infinity. Apply the following ansatz
\beeq
\Psi(t,r)=e^{-i \omega t} \psi(r),
\label{TimeDepend}
\eneq  
where  $e^{-i \omega t}$ is the time evolution of the scalar field, then we obtain the time-independent wave equation
\beeq 
\frac{d^2 \psi}{dr_*^2}+\left[\omega^2-V(r)\right]\psi=0,
\label{WEnoTime}
\eneq  
where $\omega$ is the complex QNM frequency (or eigenvalues of the above wavelike equation) written in the form $\omega=R e(\omega)+i \operatorname{Im}(\omega)$ to be determined after the appropriate boundary conditions are applied ($\psi(r)$ corresponds to the purely outgoing wave at spatial infinity; $\psi(r)$  behaves as the purely ingoing wave at the event horizon):
\begin{equation}
\psi(r)\sim \pm e^{\pm i \omega r^{*}}, \quad r^{*} \rightarrow \pm \infty.
\end{equation}

\begin{figure}
        	\centering
        	\includegraphics[width = 12 cm]{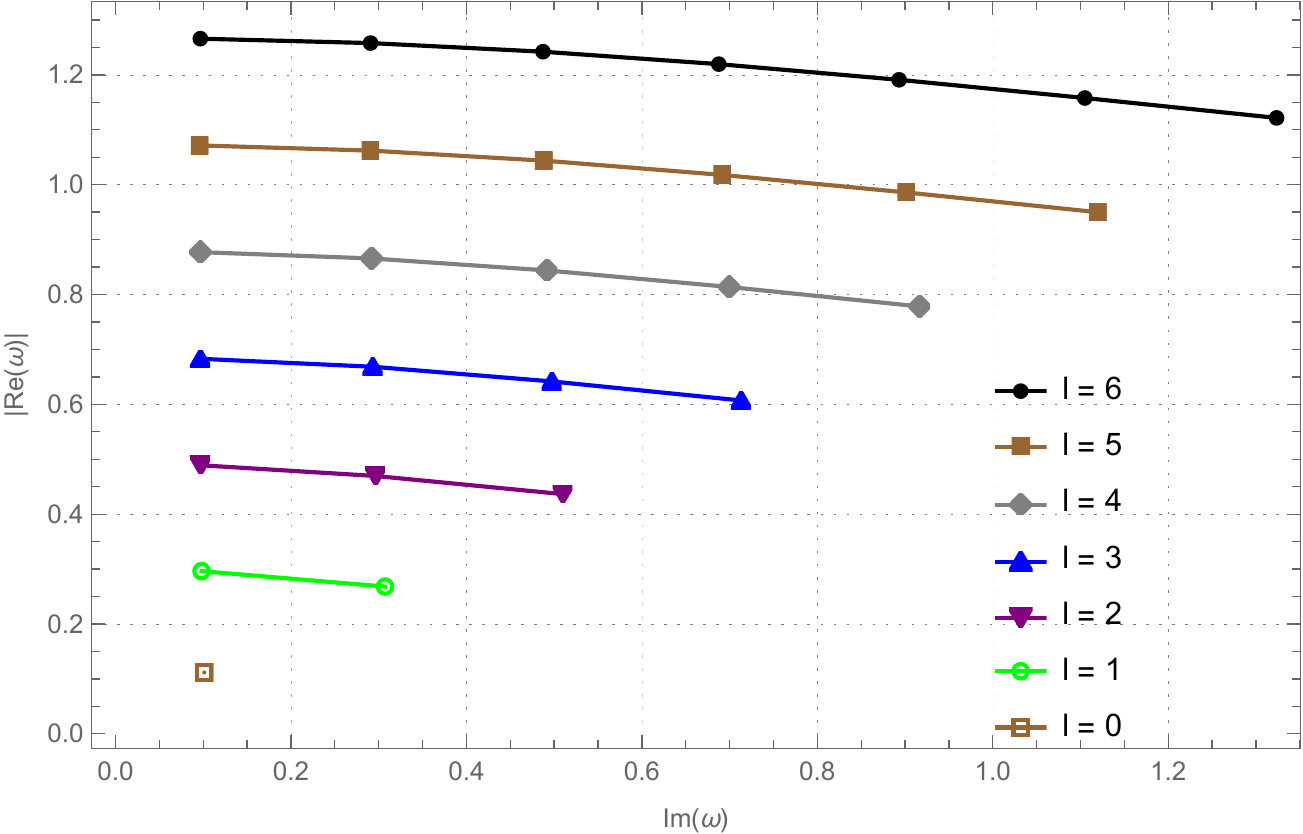}
            \caption{
            The QNM spectrum for $ \gamma = 0.2$. We show the cases in which $ l \in \{0,1,2,3,4,5,6\}$. }
        	\label{fig:QNMspectrum}
        \end{figure}

\begin{figure}[th!]
	\begin{center}
		\includegraphics[height=5.cm]{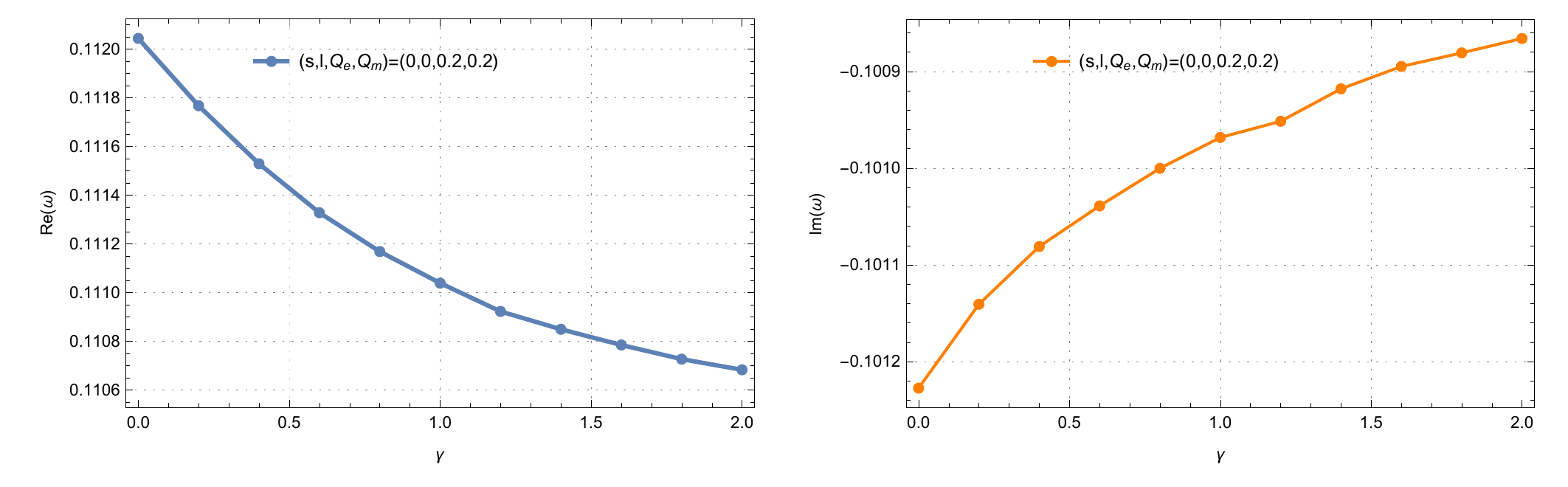}
		\includegraphics[height=5.cm]{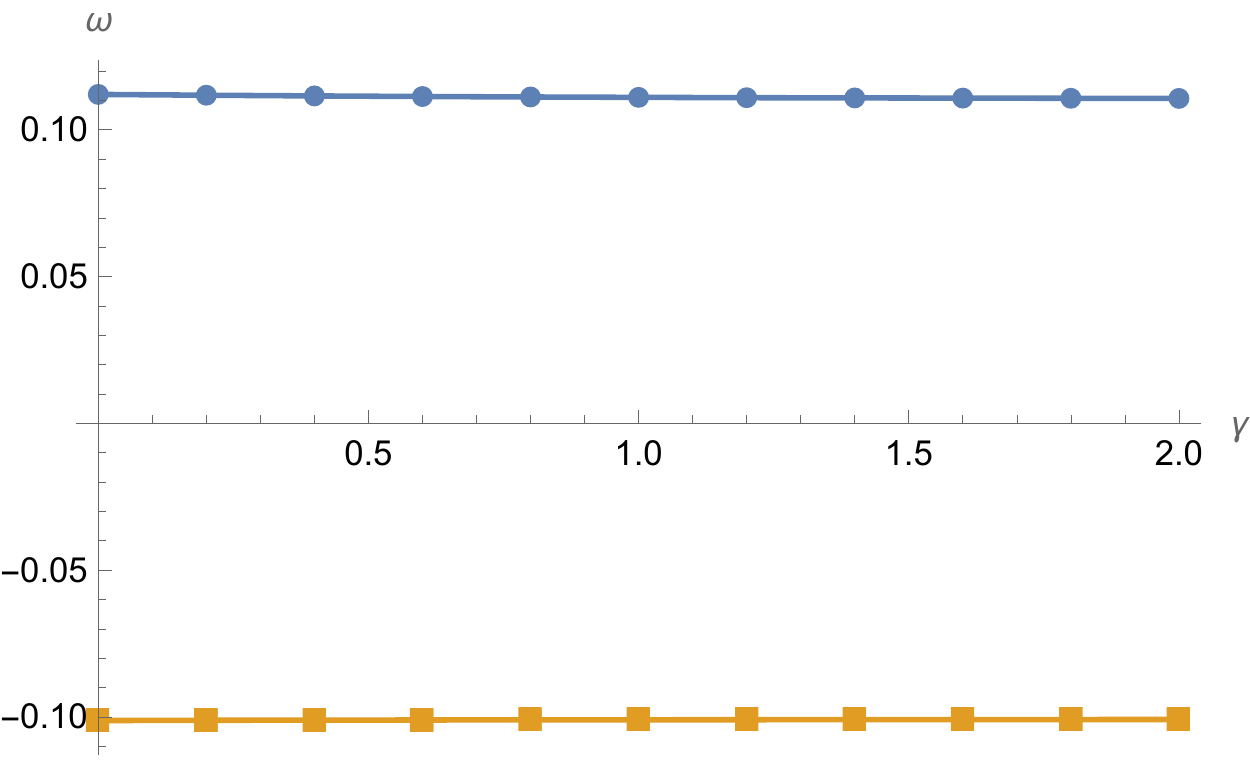}
	\end{center}
	\vspace{- 0.7cm}
	\caption{\footnotesize Scalar QNM spectrum for $s=l=0$  and $Q_\text{e}=Q_\text{m}=0.2$ for different values of $\gamma$.  }
	\label{WKB-different-l}
\end{figure}

\begin{figure}[th!]
	\begin{center}
		\includegraphics[height=5.cm]{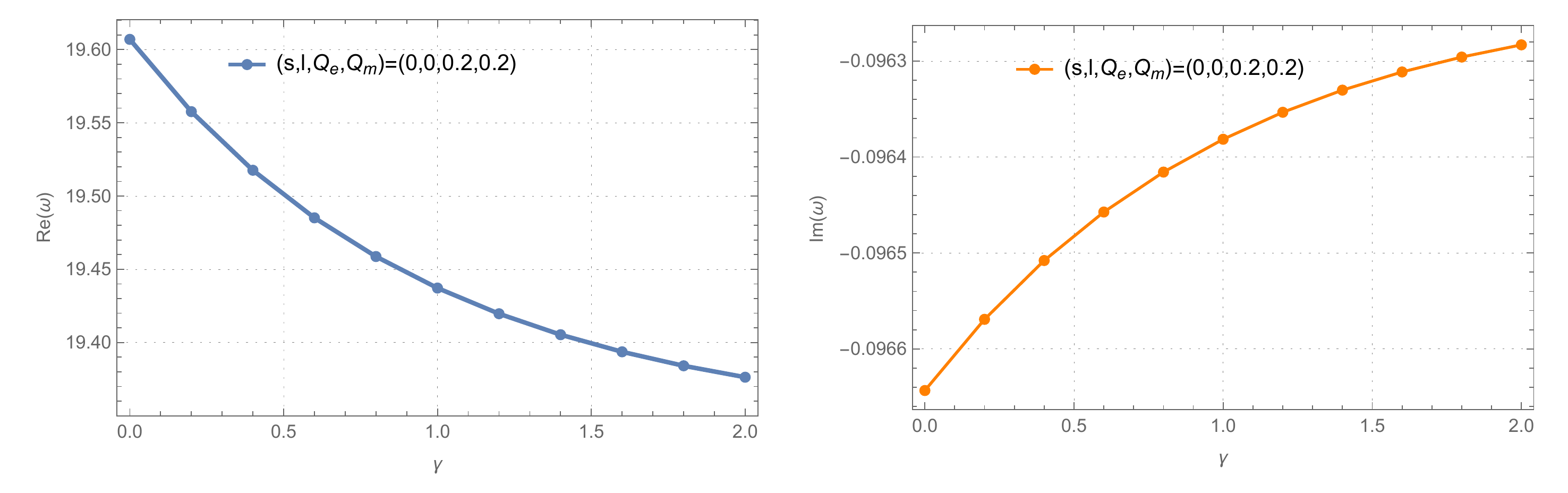}
		\includegraphics[height=5.cm]{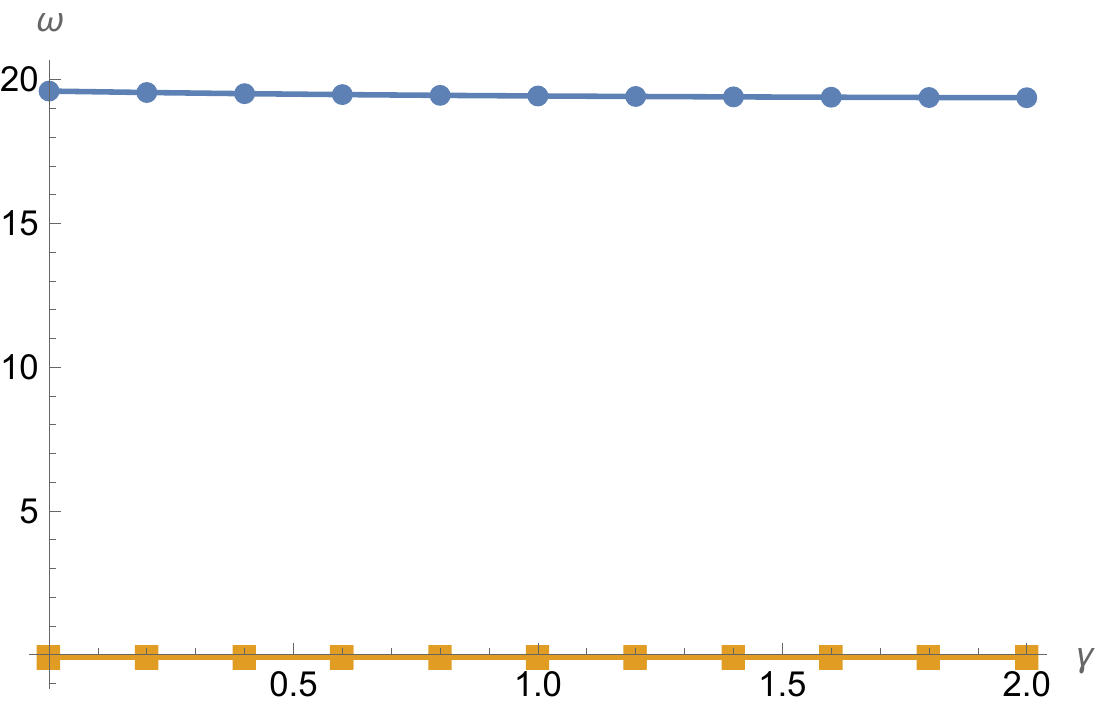}
	\end{center}
	\vspace{- 0.7cm}
	\caption{\footnotesize Scalar QNM spectrum for $s=0$, $l=100$ and $Q_\text{e}=Q_\text{m}=0.2$ for different values of $\gamma$.  }
	\label{WKB-different-l100}
\end{figure}

\begin{figure}[th!]
	\begin{center}
		\includegraphics[height=5.cm]{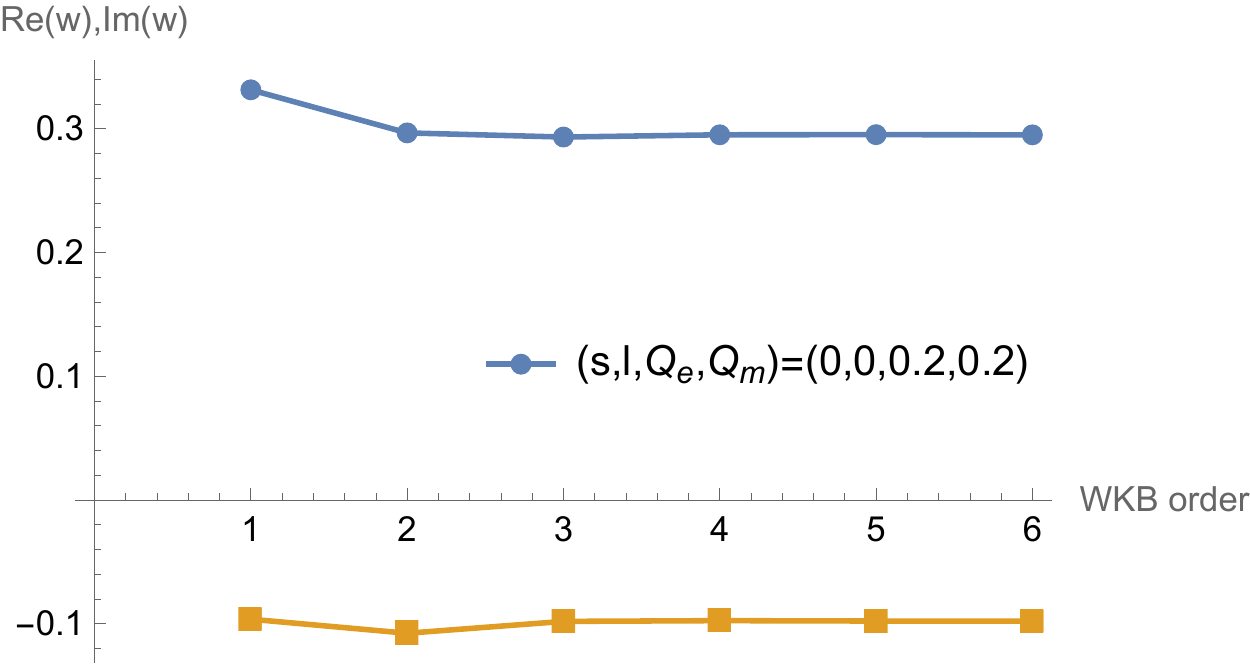}
	\end{center}
	\vspace{- 0.7cm}
	\caption{\footnotesize 
You can see the convergence of the WKB formula for $s=0$, $l=0$ and $Q_\text{e}=Q_\text{m}=0.2$ for the value of $\gamma=0.6$.  }
	\label{WKB-different-converge}
\end{figure}

Here we use the numerical method of the sixth-order WKB formula to calculate QNMS, which is based on the WKB expansion of the wave function at both the event
horizon and spatial infinity. It is then matched with the
Taylor expansion is near the peak of its effective potential. There are two turning points/monotonic decay for this potential. To do so, we should solve the following equation\cite{Iyer:1986np}
\beeq
\frac{i \left[ \omega^2 -V(r_{*})|_{\bar{r}_*}\right]}{\sqrt{2V''(r_*)|_{\bar{r}_*}}} -\overset{N}{\underset{j=2}{\sum}} \Lambda_j(n)=n+\dfrac{1}{2},
\label{WKBorder}
\eneq
where $\Lambda_j(n)$ ($N$ indicates the order of the WKB method) are the WKB correction terms, $\bar{r}_*$ is the location of the maximum of the QNM potential $V(r_*)$ in the tortoise coordinate. Note that $\Lambda_{2,3}$ are defined in \cite{Iyer:1986np}\footnote{$\Lambda_2$ in \cite{Iyer:1986np} is missing a factor of $i$ in the numerator.} and $\Lambda_{4,5,6}$ are given in \cite{Konoplya:2003ii}. The dependence of the QNM frequencies on the screening parameter $\gamma$ is
qualitatively different for lowest and higher multipoles, as seen in Table \ref{scaltablew}. We can see in Table \ref{scaltablew} and  \ref{scaltablew2} that both real and imaginary parts of the $\omega$ decrease when the parameter $\gamma$ is increased. The scalar field QNMs for different values of $l$ is plotted in \ref{fig:QNMspectrum}, as well as for $l=0,100$ for different values of $\gamma$ are presented in Figs. \ref{WKB-different-l}, \ref{WKB-different-l100} and \ref{WKB-different-converge}.

\begin{table}
    \centering
    \begin{tabular}{c c c c}
    \hline
    \hline
    \multicolumn{4}{c}{QNM via WKB $(M\omega)$ method} \\
    \hline
    $ \gamma / M$ & 3th order WKB $(M\omega)$ ($l=0$) $(M\omega)$ & 6th order WKB $(M\omega)$ ($l=0$) & 6th order WKB $(M\omega)$ ($l=100$) \\
    0.0 & 0.106364 - 0.11523$i$     & 0.112044\, - 0.101227$i$      & 19.6069 - 0.0966434 \\
    0.2     & 0.106046 - 0.115229 $i$     & 0.111767 - 0.101141$i$      & 19.5576 - 0.0965692$i$       \\ 
    0.4     & 0.105787 - 0.115226 $i$     & 0.111529 - 0.101081$i$       & 19.5176 - 0.0965079$i$       \\ 
    0.6      & 0.105577 - 0.115223$i$      & 0.111328 - 0.101039$i$      & 19.4851 - 0.0964573$i$        \\ 
    0.8      & 0.105406 - 0.115219$i$       & 0.111169 - 0.101$i$     & 19.4587 - 0.0964157$i$       \\ 
    1.0      & 0.105267 - 0.115216$i$     & 0.111039 - 0.100968$i$      & 19.4372 - 0.0963815$i$       \\ 
    1.2      & 0.105154 - 0.115213$i$     & 0.110923 - 0.100952$i$       & 19.4197 - 0.0963534$i$       \\ 
    1.4      & 0.105061 - 0.115211$i$     & 0.11085 - 0.100918$i$      & 19.4053 - 0.0963303$i$       \\ 
    1.6      & 0.104986 - 0.115208$i$     & 0.110785 - 0.100895$i$       & 19.3937 - 0.0963114$i$       \\ 
    1.8      & 0.104924 - 0.115206 $i$     & 0.110727 - 0.100881$i$      & 19.3841 - 0.0962958$i$ \\
    \hline
    \end{tabular}
    \caption{QNM frequencies of scalar field perturbation for ModMAx black hole ($M=1,Q_\text{e}=0.2,Q_\text{m}=0.2$).}
    \label{scaltablew}
\end{table}

\begin{table}
    \centering
    \begin{tabular}{c c}
    \hline
    \hline
    \multicolumn{2}{c}{QNM via WKB $(M\omega)$ method} \\
    \hline
    $\gamma/M$ & 6th order WKB $(M\omega)$ ($l=2$) $(M\omega)$ \\
    0.0 & 0.490316 - 0.097176 2$i$     \\ 
    0.2 & 0.489077 - 0.0971035 $i$  \\ 
    0.4 & 0.488072 - 0.0970434 $i$      \\ 
    0.6 & 0.487256 - 0.0969938 $i$       \\ 
    0.8 & 0.486592 - 0.096953 $i$       \\ 
    1.0 & 0.486051 - 0.0969194 $i$       \\ 
    1.2 & 0.485611 - 0.0968918 $i$         \\ 
    1.4 & 0.485251 - 0.0968692 $i$         \\ 
    1.6 & 0.484958 - 0.0968506 $i$          \\ 
    1.8 & 0.484718 - 0.0968353 $i$        \\
    \hline
    \end{tabular}
    \caption{QNM frequencies of scalar field perturbation for ModMax black hole ($M=1,Q_\text{e}=0.2,Q_\text{m}=0.2$).}
    \label{scaltablew2}
\end{table}



Next, we study the eikonal quasinormal modes for dyonic ModMax black holes. This method is also known as the geometric optics method due to its relation with the parameters on the null geodesics \cite{Cardoso:2008bp,Konoplya:2017wot}. The imaginary part of the quasinormal mode frequency (Im $\omega$=-$\omega_I$), which is responsible for the temporal, exponential decay, can be calculated in the large-$l$ limit ($l\to \infty$)
(only the $g_{tt}$ component is relevant) \cite{Glampedakis:2019dqh,Churilova:2019jqx} i.e., as the angular momentum number describing our mode solution becomes very large, as follows:
\begin{equation}
    \omega_{l \gg 1}=l \Omega_\text{ph}-i\left(n+\frac{1}{2}\right)\left|\lambda_{\mathrm{L}}\right|,
\end{equation}
with the angular velocity $\Omega_\text{ph}$:
\begin{equation}
    \Omega_\text{ph}= \frac{\sqrt{-g_{t t}\left(r_\text{ph}\right)}}{r_\text{ph}}=\frac{\sqrt{f\left(r_\text{ph}\right)}}{r_\text{ph}},
\end{equation} 
and Lyapunov exponent $\lambda_{\mathrm{L}}$:
\begin{equation}
    \lambda_{\mathrm{L}}=\sqrt{\frac{f\left(r_\text{ph}\right)\left[2 f\left(r_\text{ph}\right)-r_\text{ph}^{2} f^{\prime \prime}\left(r_\text{ph}\right)\right]}{2 r_\text{ph}^{2}}}, \label{Lyapunov}
\end{equation}
where $n$ is the overtone number and take values $n= 0, 1, 2,...$. Note that the eikonal limit is independent of the spin of the perturbation, so that black holes' scalar, electromagnetic, and gravitational perturbations give the same behaviour in the eikonal limit \cite{Kodama:2003jz}. Table \ref{tab:table2} shows that the real parts decrease, but the imaginary part of the QNMs increase with the increasing parameter $\gamma$. We can conclude that these modes are stable because the imaginary parts of the QNMs frequencies are negative. The decay rates (imaginary part) of QNMs frequencies increase as the screening parameter $\gamma$ increases.
\begin{table}[ht!]
    \centering
    \begin{tabular}{ c c c }
    \hline
    \hline
    $\gamma/M$ &  $\omega_{R}$ & $\omega_{I}$ \\
    \hline
    0.2 & 19.4602  & 0.862748   \\
    0.6 & 19.3881 & 0.863857  \\
    0.1 & 19.3404  & 0.864585   \\
    1.4 & 19.3087 &  0.865065   \\
    1.8 & 19.2876  & 0.865384   \\
    \hline
    \end{tabular}
    \caption{Effects of the parameter on the frequencies of the quasinormal modes in eikonal limits for fixed  $Q_\text{e}=Q_\text{m}=0.2M$, $s=1$, $n=0$, $l=100$.}
    \label{tab:table2}
\end{table}

 To study time-domain profiles of the scalar field perturbation of the BHs, we use the initial disturbance as a Gaussian wave packet:
        \begin{equation}
            \Psi\left(r_{*}, t = 0\right)=\mathcal{A}\, \text{Exp} \left(-\frac{\left(r_{*}-\bar{r}_{*}\right)^{2}}{2 \sigma^{2}}\right),\left.\partial_{t} \Psi(r_*, t)\right|_{t=0}=-\partial_{r_{*}} \Psi\left(r_{*}, 0\right).
        \end{equation}
        
\begin{figure}
        	\centering
        	\includegraphics[width = 8 cm]{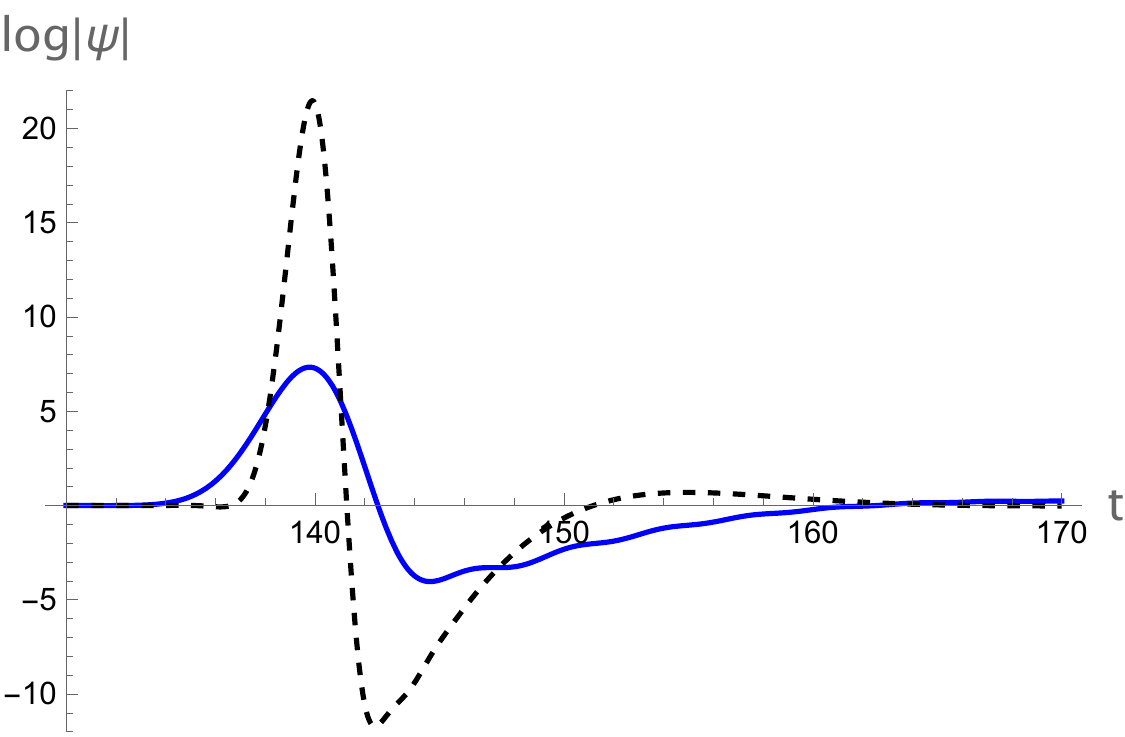}
        	\includegraphics[width = 8 cm]{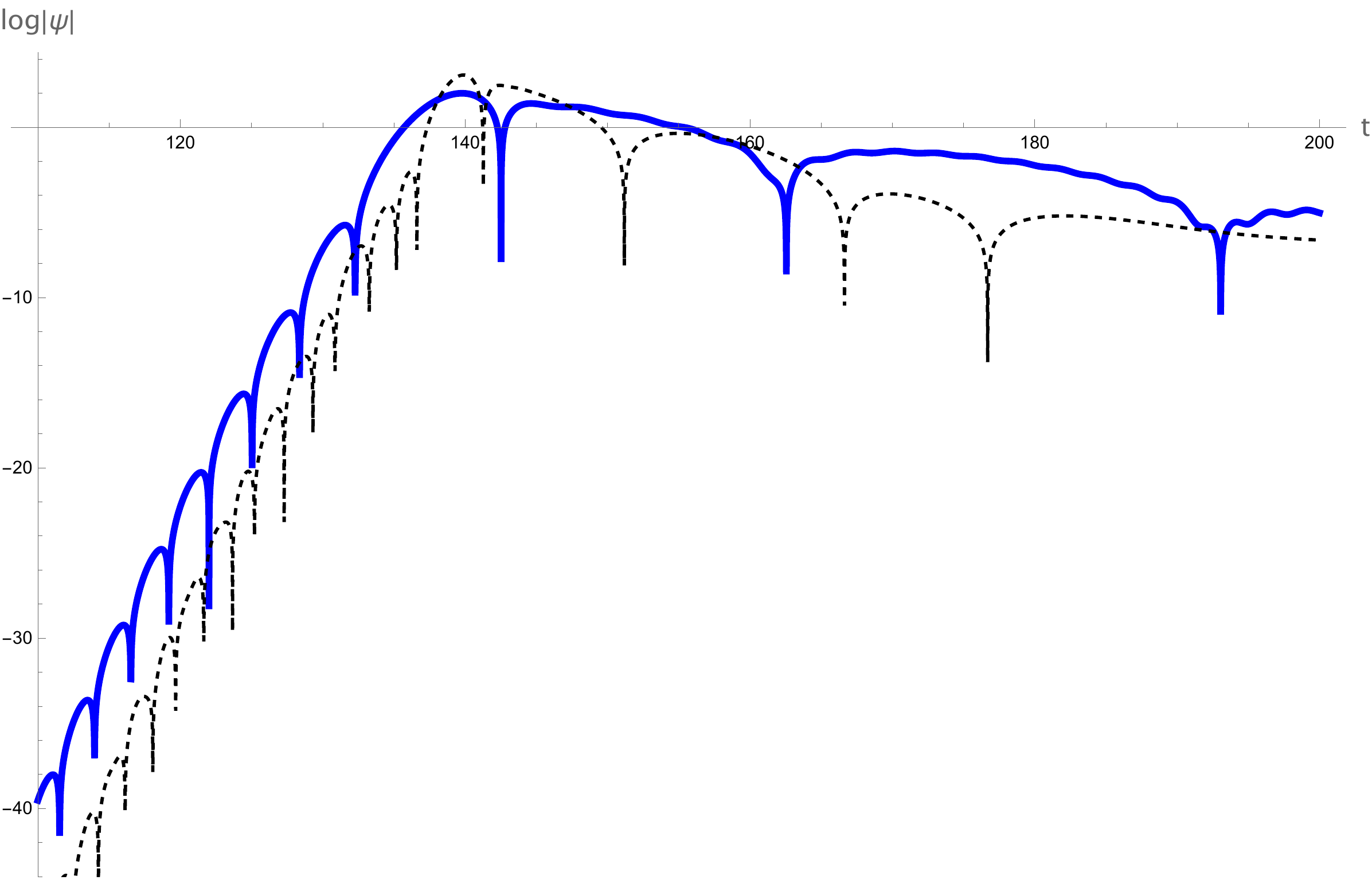}
            \caption{
           The time-domain profiles of the scalar field perturbation for the case l = 0. The ModMax black hole is given in blue while the Schwarzschild case is in black. In the logarithmic plot, we see that the ModMax frequency is slightly lower.
                }
        	\label{fig:waveforml2}
        \end{figure}

\begin{figure}
        	\centering
        	\includegraphics[width = 8 cm]{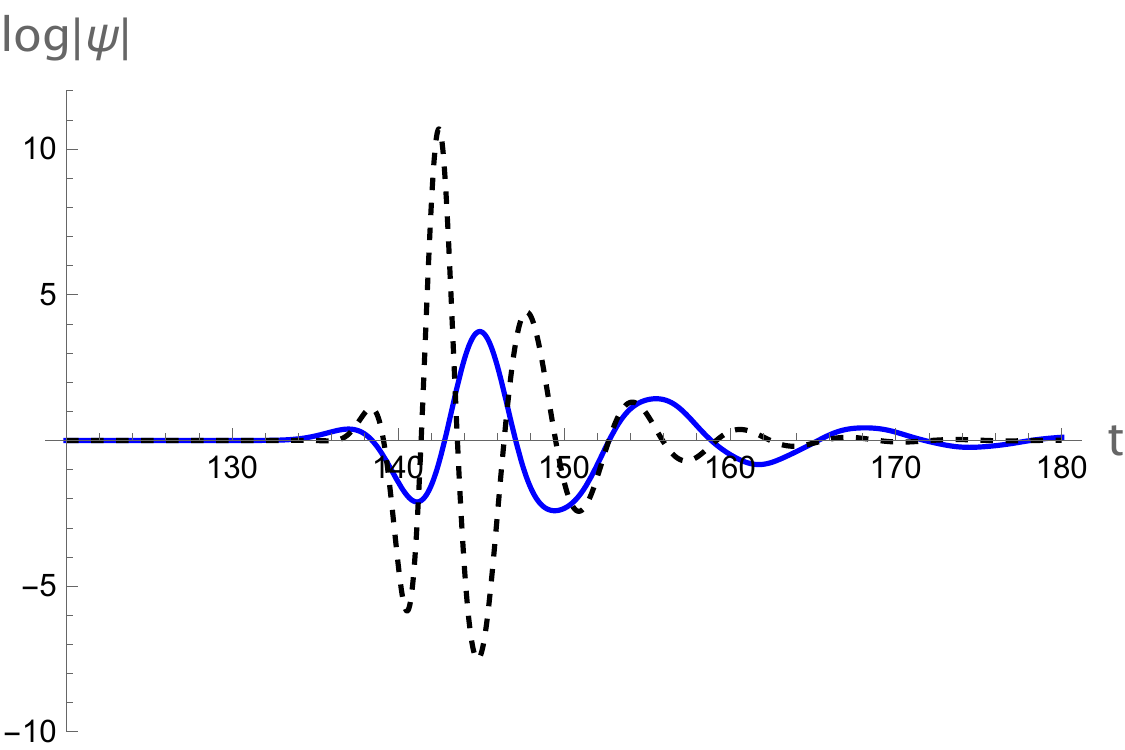}
        	\includegraphics[width = 8 cm]{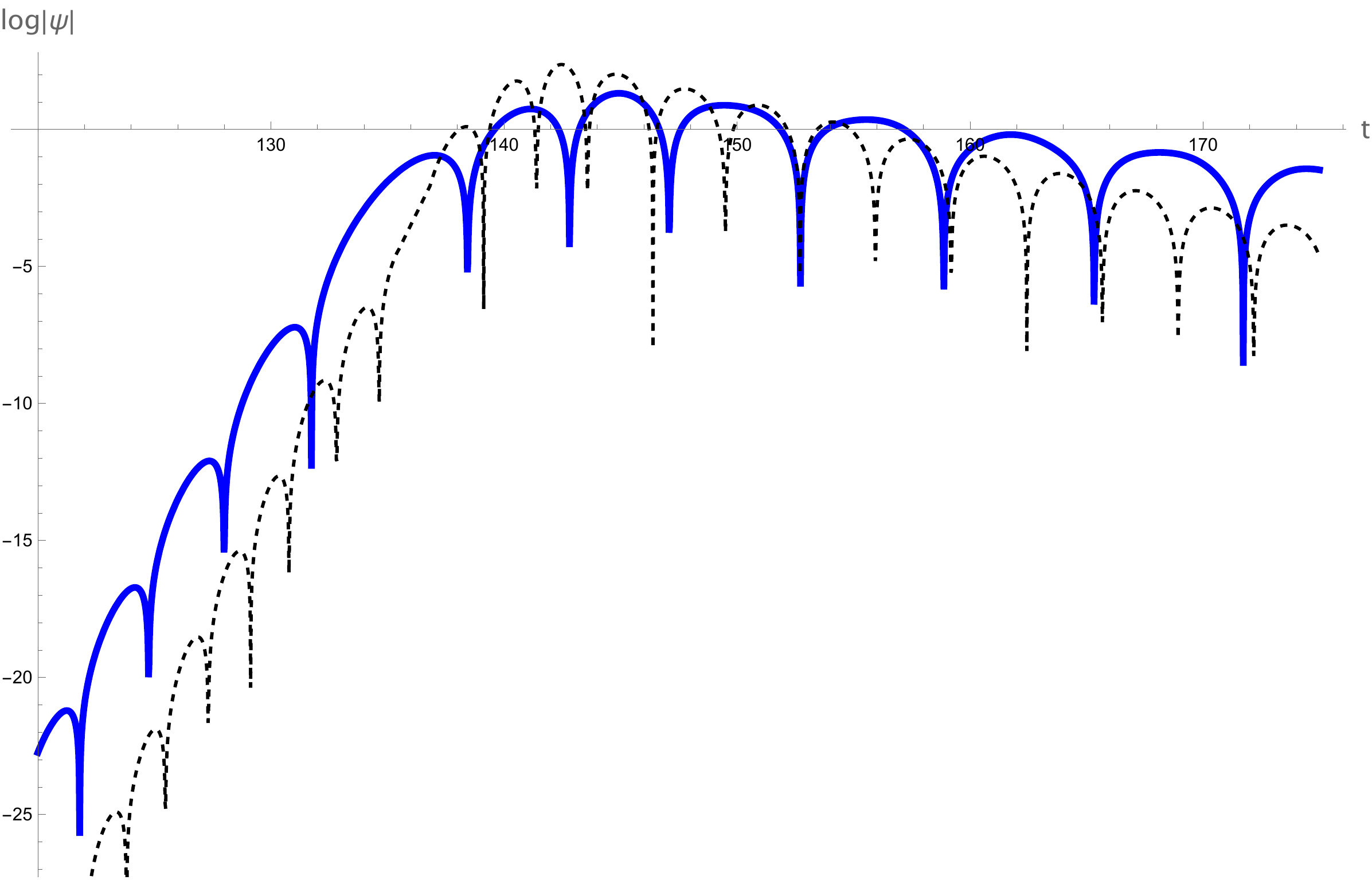}
            \caption{
           The time-domain profiles of the scalar field perturbation for the l = 2 cases. The ModMax black hole is given in blue, while the Schwarzschild case is in black. The logarithmic plot shows that the ModMax frequency is slightly lower.
                }
        	\label{fig:waveforml3}
        \end{figure}

It is noted that we assume that $\sigma=2, \bar{r}_{*}=-40$, and $\mathcal{A}=10$. Then apply the boundary conditions such that the wavefunction vanishes at ($r_* = -200, r_* = 250$) to observe the differences between the Schwarzschild black hole and ModMax black hole spacetimes. Afterwards, the time-domain profiles of the scalar field perturbation for $l=0$ and $l=2$ cases are plotted numerically in Figs. \ref{fig:waveforml2} and \ref{fig:waveforml3}. The ModMax black hole is given in blue, while the Schwarzschild case is in black. The logarithmic plot shows that the ModMax frequency is slightly lower. For the $l=0$ case, the plot shows the relatively short period of quasinormal ringing, which makes it hard to extract values of frequencies with good accuracy.

To examine the gravitational signal emitted by an oscillating black hole within the bandwidth of the LIGO/VIRGO interferometers, its frequency must be in the range of around  $10-40$ Hz, or for the LISA interferometer, about $10^{-4}-10^{-1}$ Hz \cite{Ferrari:2007dd}. We assume  $M=\alpha M_{\odot}, \,\left(M_{\odot}=1.48 \times  10^{5}\, \mathrm{~cm}\right)$, for the frequencies and damping times obtained by \cite{Ferrari:2007dd}

        \begin{equation}
        \nu=\frac{c}{2 \pi \alpha \times  M_{\odot}\left(M \omega_{0}\right)}=\frac{32.26}{\alpha}\left(M \omega_{0}\right) \mathrm{kHz}, \quad \tau=\frac{\alpha M_{\odot}}{\left(M \omega_{i}\right) c}=\frac{\alpha \times  0.4937 \times  10^{-5}}{\left(M \omega_{i}\right)} \mathrm{s}.
        \end{equation}
        
After calculating the fundamental frequencies, we show the allowed range of masses of oscillating ModMax  black holes which LIGO and LISA can detect:
        \begin{equation}
            13 M_{\odot} \lesssim M \lesssim 1.3 \times 10^{3} M_{\odot},
        \end{equation}
        
for the fundamental frequency range $\nu \in[12 \mathrm{~Hz}, 1.2 \mathrm{kHz}]$, and for LISA the detectable mass range is 
        
        \begin{equation}
             1.56\times 10^{5} M_{\odot} \lesssim M \lesssim 1.56 \times  10^{8} M_{\odot}.
        \end{equation}
        
Note that it corresponds to frequencies $\nu \in\left[10^{-4}, 10^{-1}\right] \mathrm{Hz}$. Moreover,  in the future, LISA will be able to detect signals emitted by the oscillations
of the massive black hole at the centre of our Galaxy SGR A*, the mass of
which is $M=4.1 \times 10^6M_{\odot} $ \cite{EventHorizonTelescope:2022xnr}.\\

\section{Bounds of Greybody factors and High-Energy Absorption cross-section} \label{sec6} 

Greybody factor (GF) is a quantity related to the quantum nature of a black hole, and its high value gives a high probability that Hawking radiation can reach infinity, so the GF of test fields is important for estimating the intensity of Hawking radiation. Here, we study the rigorous bound on the GF of the dyonic ModMax black hole to probe the effect of $\gamma$ on the bound. Rigorous bound of the GF was first proposed in \cite{Visser:1998ke,Boonserm:2008zg} which gives qualitative
description of a black hole as follows:

\begin{equation}
T \geq \operatorname{sech}^{2}\left(\int_{-\infty}^{\infty} \vartheta d r_{*}\right),
\end{equation}
with 
\begin{equation}
\vartheta=\frac{\sqrt{\left[h^{\prime}\left(r_{*}\right)\right]^{2}+\left[w^{2}-V\left(r_{*}\right)-h^{2}\left(r_{*}\right)\right]^{2}}}{2 h\left(r_{*}\right)}.
\end{equation}

Note that  $h\left(r_{*}\right)$ is a  function satisfying the condition $h(-\infty)=$ $h(\infty)=w .$ \cite{Visser:1998ke}.  Selecting $h=w$, and substituting the tortoise coordinate  $r_*$, we write
\begin{equation} \label{bound}
T_{b} \geq \operatorname{sech}^{2}\left(\frac{1}{2 w} \int_{-\infty}^{\infty}\left|V\right| \frac{d r}{f(r)} \right).
\end{equation}

\begin{figure}
    \centering
    \includegraphics[width=0.48\textwidth]{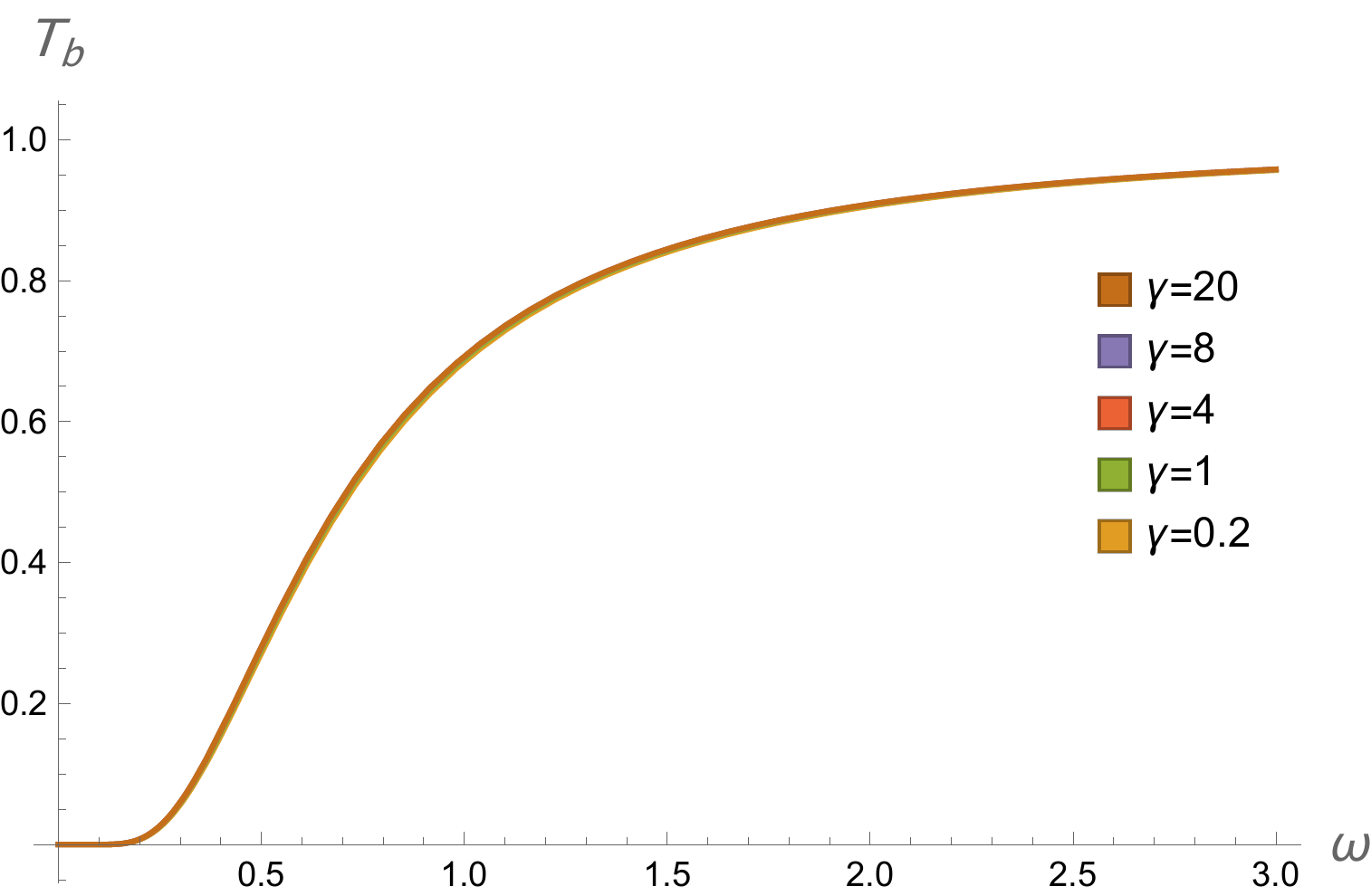}
    \includegraphics[width=0.48\textwidth]{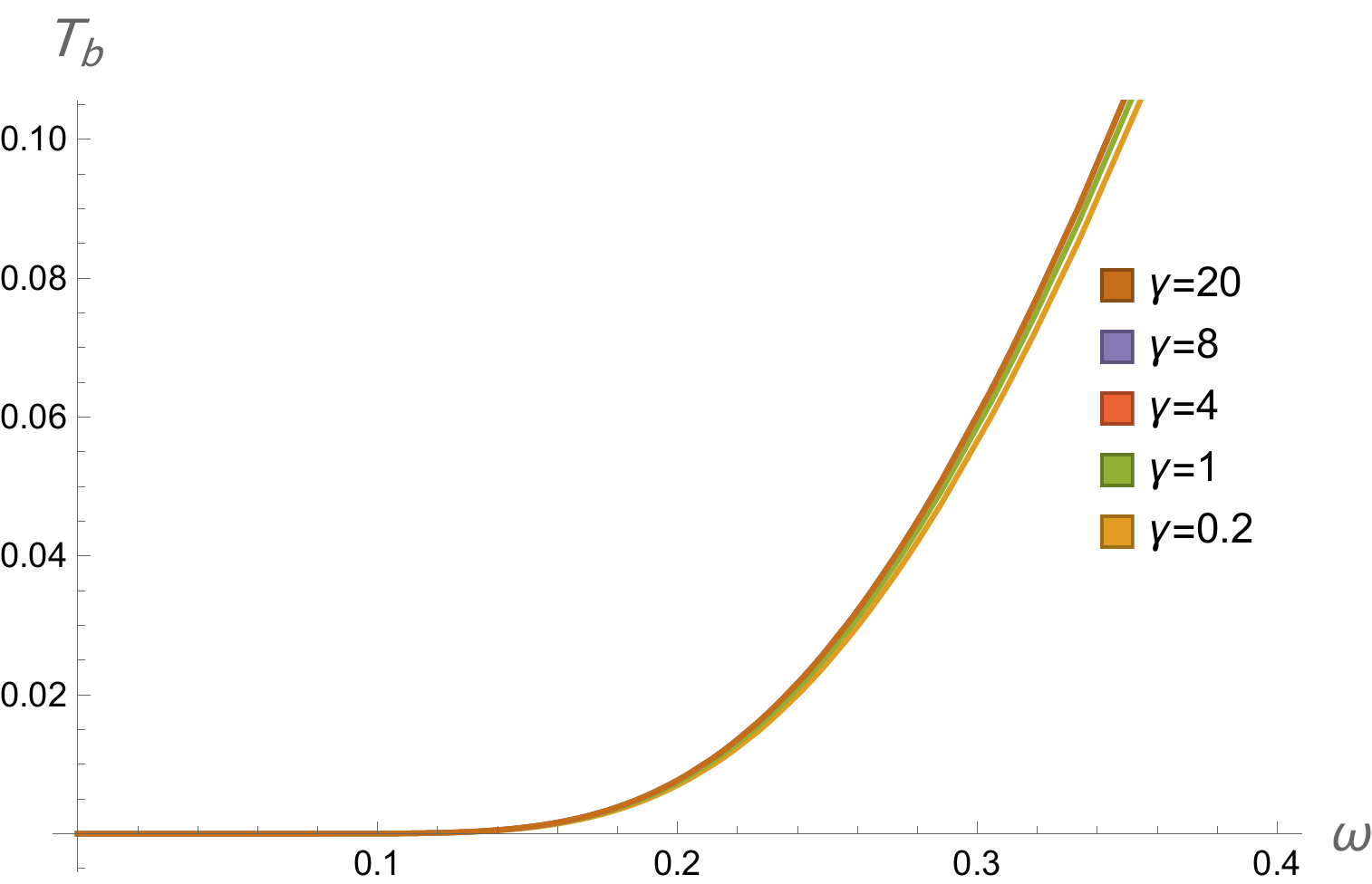}
    \caption{(Scalar Field) The Greybody Bound $T_b$ versus the $w$ for different values of values of $\gamma$ parameter, with $M = 1$, $l=1$, $s=0$, and $Q_\text{e} = Q_\text{m}=0.12m$.}
    \label{fig:greybody}
\end{figure}

Here $V$ is the effective potential for the massless scalar field given in (\ref{scalarV}). We can calculate the bound as follows:
\begin{equation}
    T \geq T_b = \text{sech}^2\left(\frac{\frac{l (l+1)}{e^{-\frac{\gamma}{2}} \sqrt{-Q_\text{e}^2+e^\gamma M^2-Q_\text{m}^2}+M}+\frac{M}{\left(e^{-\frac{\gamma}{2}} \sqrt{-Q_\text{e}^2+e^\gamma M^2-Q_\text{m}^2}+M\right)^2}-\frac{e^{-\gamma} \left(2 Q_\text{m}^2+2 Q_\text{m}^2\right)}{3 \left(e^{-\frac{\gamma}{2}} \sqrt{-Q_\text{e}^2+e^\gamma M^2-Q_\text{m}^2}+M\right)^3}}{2 w }\right).
\end{equation}
The bound reduces to the Schwarzschild case for $ (\gamma,Q_\text{e},Q_\text{m})\rightarrow 0$, as $T_\text{Sch} \geq\text{sech}^2\left(\frac{2 l (l+1)+1}{8 m w }\right)$. We illustrate the effect of the screening parameter $\gamma$ 
on the greybody bound for a scalar field in the ModMax black hole in Fig. \ref{fig:greybody}. It is seen that when the value of $\gamma$ parameter increases, the greybody bound $T_b$ grows larger.

Afterwards, the high-energy absorption cross-section is studied by applying the Sinc approximation, which Sanchez first studied for the Schwarzschild black hole. It is shown that increasing the frequency for the ordinary material sphere monotonically increases the absorption cross-section oscillated around the constant geometric-optics value for the black hole (related to the photon sphere) \cite{Sanchez:1977si} which shows the relation between the impact parameter and cross-section of the photon sphere at the critical value. 

On the other hand, it was concluded that at low energy scales, the characteristic properties of BHs and the cross-section of BHs equal the BH area \cite{Das:1996we}. However, at high energies, one should use the complex angular momentum technique to study the geometrical cross-section of the photon sphere \cite{Decanini:2011xi}. Decanini et al. use the Regge pole techniques to show the relationship between the oscillatory pattern of the high-energy absorption cross-section and the Sinc(x) function (with $\operatorname{Sinc}(x)$ denote the sine cardinal
$\operatorname{sinc}(x) \equiv \frac{\sin x}{x}
$) with the photon sphere. This method extensively discussed in the literature \cite{Magalhaes:2020sea,Paula:2020yfr,Lima:2020seq,Boonserm:2019mon,Xavier:2021sje,Richarte:2021fbi,Li:2022wzi}.

\begin{figure}
	\includegraphics[width=0.48\textwidth]{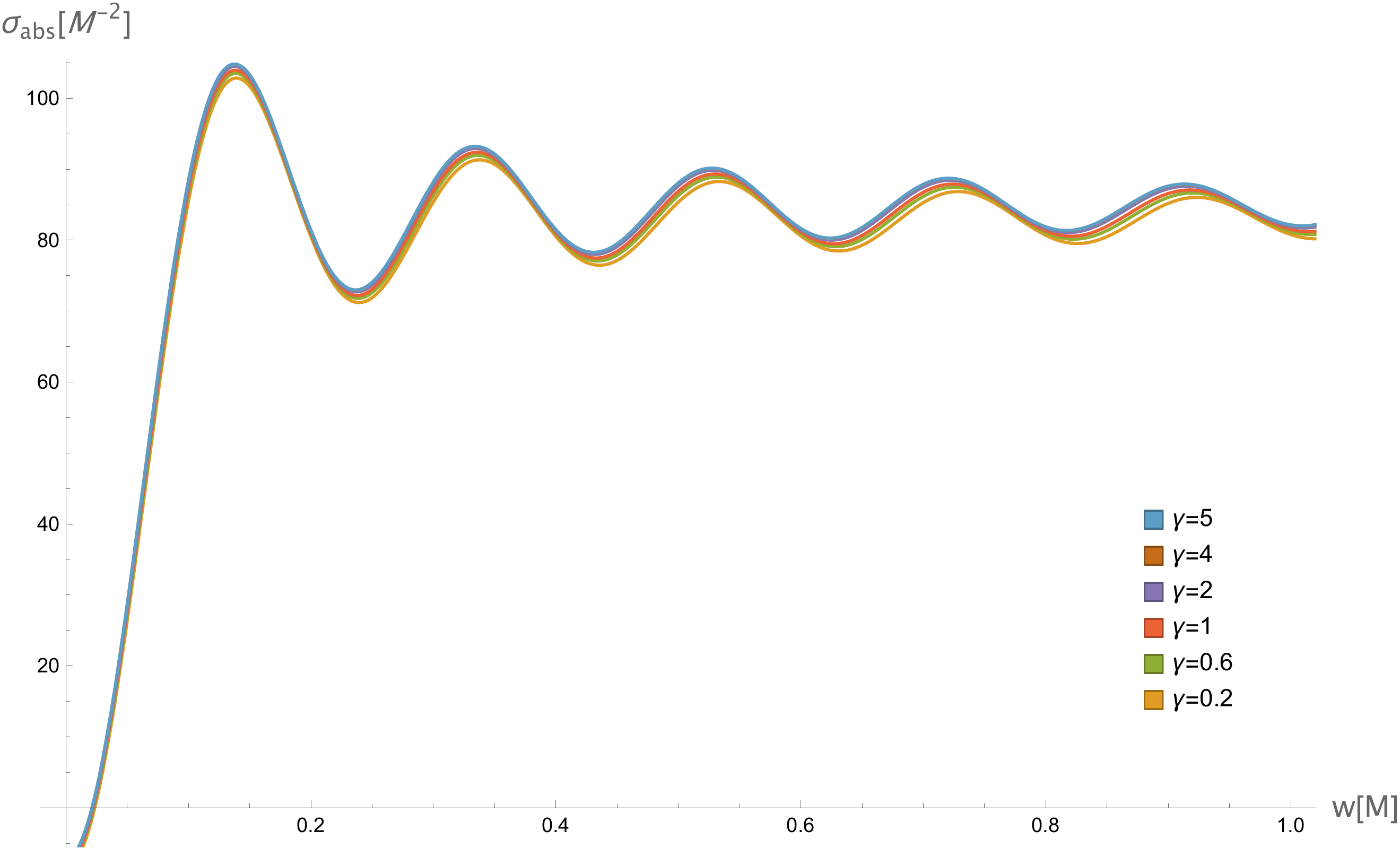}
    \caption{The total absorption cross section for various values of the $\gamma$, parameter with $M=1$, and $Q_\text{e}=Q_\text{m}= 0.2m$.}
    \label{fig:totabsorption}
\end{figure}

In the limit of high energy, the oscillatory part of the absorption cross-section can be calculated by using
 \cite{Decanini:2011xi}:
 
 \begin{equation}
\sigma_{\mathrm{abs}}^{\mathrm{osc}}(w)=-8 \pi \sigma_{\text {geo }} n_c e^{-\pi n_c} \operatorname{sinc}\left[2 \pi b_\text{crit} w\right]
\end{equation}
where the $n_{c}=\left[\left(f_{r_\text{ph}}-\frac{r_\text{ph}^{2}}{2} f_{r_\text{ph}}^{\prime \prime}\right)\right]^{1 / 2}$, the eikonal cross-section is $\sigma_{\text {geo }}=\pi b_\text{crit}^{2}$,  with the critical impact parameter given in (\ref{ebcrit}) as  $ b_\text{crit}=\frac{r_\text{ph}}{\sqrt{f\left(r_\text{ph}\right)}}$ and Lyapunov exponent  $\lambda_L$ is given in Eq. (\eqref{Lyapunov}). The absorption cross-section can also be written as
 \begin{equation}
\sigma_{\mathrm{abs}}^{\mathrm{osc}}(w)=-4 \pi \frac{\lambda_L b_\text{crit}^{2}}{w} e^{-\pi \lambda_L b_\text{crit}} \sin \frac{2 \pi w}{\Omega_{crit}}
\end{equation}

where $\Omega_\text{ph}$ is the angular velocity with the radius of the photon sphere $r_\text{ph}$. The total high energy formula for the absorption
cross-section is equal to $\sigma_\text{abs} \approx \sigma_{\mathrm{abs}}^{\mathrm{osc}} + \sigma_\text{geo}$ \cite{Decanini:2011xi,Decanini:2009mu}. In Fig. \ref{fig:totabsorption} the total absorption cross section for various values of $\gamma$ is plotted. The numerical analysis
shows that the greater values of screening parameter $\gamma$, the total absorption cross section becomes more and more unstable for higher energies and exhibits the largest amplitude. Moreover, there is a regular oscillatory behaviour around the high-frequency limit. Fig. \ref{fig:totabsorption} presents the absorption spectrum as a function of the frequency, making clear oscillations characteristic of a diffraction pattern, where the oscillations occur around its constant value of the geometrical optics with decreasing amplitude and constant period. 

\section{Effects on neutrino} \label{sec7}
\subsection{Flavour oscillation}
\begin{figure}
    \centering
    \includegraphics{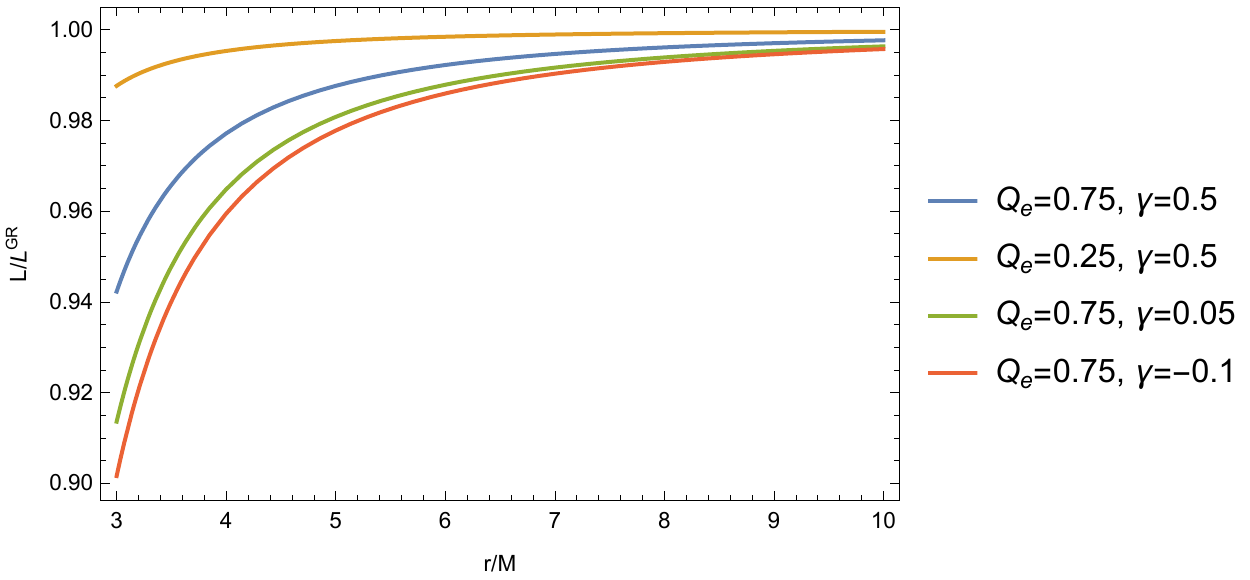}
    \caption{Neutrino oscillation length in the metric in Eq.~(\ref{eq:fdyon}) over the same quantity in GR for various parameters set $(Q_\text{m},\gamma)$ reported in the legend. For all the curves, we fixed $Q_\text{m}=0.25$.}
    \label{fig:Losc}
\end{figure}

In this Section we will study the effects of the metric in Eq.~(\ref{eq:fdyon}) on neutrinos, following the treatment in Ref.~\cite{Cardall:1996cd,Fornengo:1997qu,MosqueraCuesta:2017iln,Mastrototaro:2021kmw}. Even if in the Standard Model neutrinos are assumed massless, in the last fifty years there have been various detection of the \textit{neutrino flavour oscillation} phenomenon. The common explanation for the latter is to take into account the existence of small neutrino masses and that neutrino mass and flavour eigenstates are not coincident. Therefore, neutrinos will be treated as massive particles in this Section. Firstly, let us define the oscillation length:
\begin{equation}
    L_{\rm osc}=2\pi \, \frac{dl_{\rm pr}}{d\Phi_{\alpha\beta}} \,\ ,
\end{equation}
where $dl^2_{\rm pr}=-g_{ij}dx^idx^j$ is the infinitesimal proper distance, with $i,j$ that runs over the spatial coordinates, $\Phi_{\alpha\beta}=\Phi_{\alpha}-\Phi_{\beta}$ and:
\begin{equation}
    \Phi_{\alpha}=\int dr\frac{m_{\alpha}}{\dot{r}}=\int dr\frac{m_{\alpha}^2}{\sqrt{E^2-g_{00}(r)\left(\frac{L^2}{r^2}-m^2_{\alpha}\right)}} \,\ ,
\end{equation}
where $L$ is the angular neutrino momentum. Considering $E\gg m^2_{\alpha}$ and $L=0$, one obtains
\begin{equation}
    L_{\rm osc}=\frac{2\pi E}{\sqrt{g_{00}}(m^2_{\alpha}-m^2_{\beta})} \,\ .
    \label{eq:Losc}
\end{equation}
Using Eq.~(\ref{eq:Losc}), we show in Fig.~\ref{fig:Losc} the behaviour of the oscillation length in $r$ for the different parameters of the metric in Eq.~(\ref{eq:fdyon}). As it is possible to see, the main differences in General relativity (GR) occur at low values or $r$, up to a $10$\% of changes in the neutrino oscillation length.

\begin{figure}
    \centering
    \includegraphics{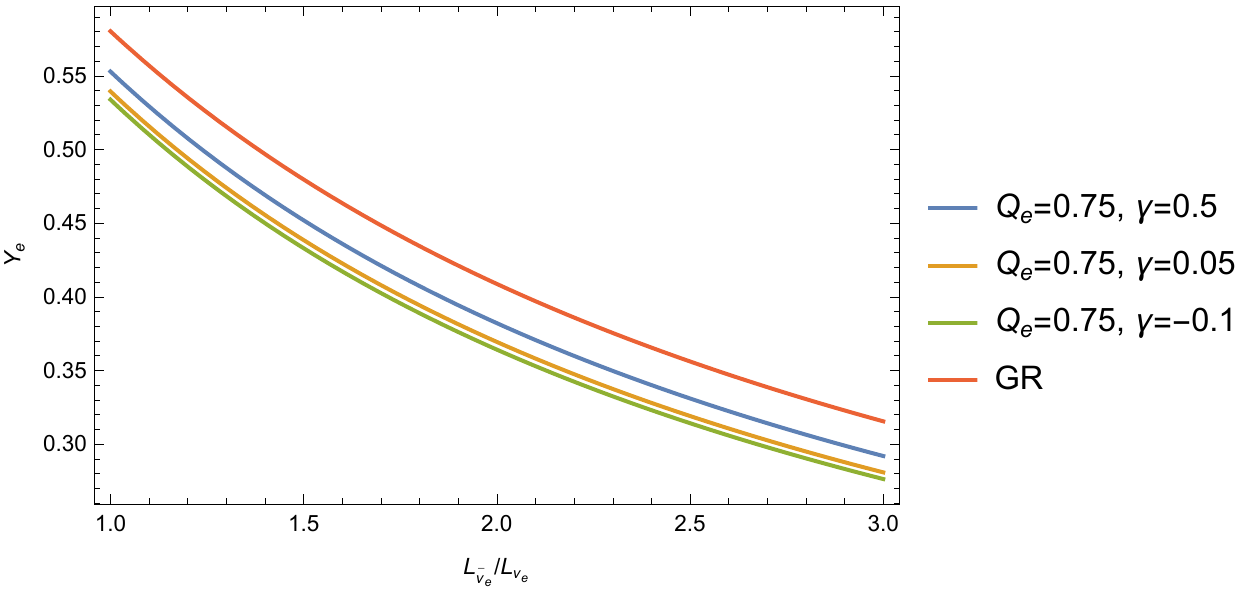}
    \caption{$Y_e$ parameter for the metric in Eq.~(\ref{eq:fdyon}) for various parameters set $(Q_\text{e},\gamma)$ reported in the legend. For all the curves, we fixed $Q_\text{m}=0.25$.}
    \label{fig:Ye}
\end{figure}

Moreover, we can discuss the effects of gravitational redshift of the considered metric on the energy spectra of $\nu_e$ and $\bar{\nu}_e$ Type II supernova explosion. The gravitational effects affect the r-process nucleosynthesis in astrophysical environments, and we might theoretically estimate that with the $Y_e$ parameter~\cite{MosqueraCuesta:2017iln,Mastrototaro:2021kmw,Fuller:1996kt,Wanajo_2014}:
\begin{equation}
    Y_e=\frac{1}{1+R_{n/p}}\,\ , \quad\quad R_{n/p}=R_{n/p}^0\Gamma \,\ ,
\end{equation}
with the local neutron to proton ratio $R_{n/p}^0$ and $\Gamma$ are defined as:
\begin{equation}
    R_{n/p}^0\simeq\left[\frac{L_{\bar{\nu}_e}\langle E_{\bar{\nu}_e}\rangle}{L_{\nu_e}\langle E_{\nu_e}\rangle}\right]\,\ , \quad\quad \Gamma=\left[\frac{g_{00}(r_{\bar{\nu}_e})}{g_{00}(r_{\nu_e})}\right]^{\frac{3}{2}} \,\ ,
\end{equation}
where $L_{\bar{\nu}_e}$ and $L_{\nu_e}=$ are the  antineutrinos and neutrinos luminosity respectively, $\langle E_{\bar{\nu}_e}\rangle=25~\mathrm{MeV}$, $\langle E_{\nu_e}\rangle=10~\mathrm{MeV}$ and $r_{\bar{\nu}_e}=5~\mathrm{km},r_{\nu_e}=0.9r_{\bar{\nu}_e}$ are the two neutrino-sphere radius for $\bar{\nu}_e$ and $\nu_e$ respectively. In Fig.~\ref{fig:Ye}, we show the behaviour of $Y_e$ for the different ratio of the $\nu_e,\bar{\nu}_e$ luminosity with the metric in Eq.~(\ref{eq:fdyon}). It is possible to see that, compared to GR (red curve), we obtain a difference up to $5$\%, which could be important in the supernovae' evolution. 

\subsection{Spin oscillation}
In this subsection, we treat the neutrino spin oscillation problem in the metric in Eq.~\eqref{eq:fdyon}, following the Ref.~\cite{Mastrototaro:2021kmw,Dvornikov:2020oay,Obukhov:2009qs}. The evolution of the spin vector $\mathbf{s}^a$ is given by
\begin{equation}
    \frac{d\mathbf{s}^a}{dt}=2\,\bm{\zeta}\times\mathbf{\Omega_g} \,,
    \label{evolution}
\end{equation}
where $\mathbf{\zeta}$ and $s^a$ are defined as
\begin{align}
    \mathbf{s}^a&=\left(\bm{\zeta\cdot u},\mathbf{\zeta}+\frac{\bm{u(\zeta \cdot u)}}{1+u^0}\right) \,, \\
    u&=(u^0,\mathbf{u}) \,,
\end{align}
with $u$ the four-velocity in the local Minkowski space.
In the metric of our interest, $\Omega_g=(0,\Omega_2,0)$ and therefore we can use the following representation for $\bm{\zeta}=(\zeta_1,0,\zeta_3)=(\cos{\alpha},0,\sin{\alpha})$. Focusing on the neutrino spin oscillation, we define the helicity $h=\bm{\zeta \cdot u}/|\bm{u}|$. The initial condition is 
\begin{eqnarray}
h_{-\infty}&=&-1 \,\ ,\\ \bm{u}_{-\infty}&=&\left(-\frac{\sqrt{E^2-m^2}}{m},0,0\right) \,\ , \\
\bm{\zeta}_{-\infty}&=&(1,0,0) \,\ ,\\ \alpha_{-\infty}&=&0 \,\ .
\end{eqnarray}
Moreover, we can write that $\bm{u}_{+\infty}= \left(+\sqrt{E^2-m^2}/m,0,0\right)$ and therefore $h_{+\infty}=\cos{\alpha}$. From that, we write the  helicity states as
\begin{align}
    \psi_{-\infty}&=\ket{-1} \,\ \\
    \psi_{+\infty}&=a_+\ket{-1}+a_-\ket{1} \,,
\end{align}
where $a_+^2+a_-^2=1$ due to the normalization and $a_+^2-a_-^2=\cos\alpha=\langle h\rangle_{+\infty}$. Therefore, one finds that
\begin{equation}\label{PLR18}
P_{LR}=|a_-|^2=\frac{1-\cos\alpha_{+\infty}}{2} \,.
\end{equation}
Using Eq. (\ref{evolution}), we obtain
\begin{eqnarray}
    \frac{d\sin\alpha}{dt}&=&2\cos\alpha\Omega_2\quad\rightarrow\quad\alpha=2\Omega_2 t \,.\\
    \frac{d\alpha}{dr}&=&\frac{d\alpha}{dt}\frac{dt}{dr}=\frac{d\alpha}{dt}\frac{dt}{d\tau}\frac{d\tau}{dr} \,,
\end{eqnarray}
where $dt/d\tau=U^0$ and $dr/d\tau=U^1$. Finally, the angle $\alpha_{+\infty}$ reads
\begin{equation}
    \alpha_{+\infty}=\int dr\frac{d\alpha}{dr} \,\ .
    \label{21}
\end{equation}

\begin{figure}
    \centering
    \includegraphics{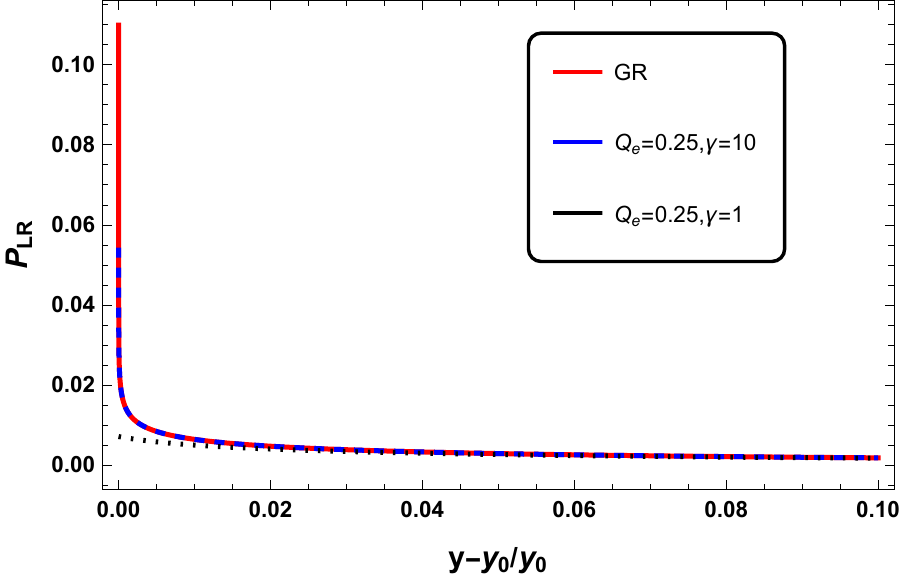}
    \caption{Spin flip probability for the metric in Eq.~\eqref{eq:fdyon}, for different parameter values specified in the legend (we set always $Q_\text{m}=0.25$). We have also shown the GR spin-flip probability (red line) for comparison. We have considered neutrinos with $\gamma=E/m=10$.}
    \label{fig:spinflip}
\end{figure}

Using Eq.~\eqref{21} it is possible to obtain the spin-flip probability for metric in Eq.~\eqref{eq:fdyon}, shown in Fig.~\ref{fig:spinflip}. The important feature is that, with the used metric, the spin-flip phenomenon is suppressed. The analysis with the inclusion of a magnetic field will be faced elsewhere.

\section{Conclusions} \label{sec8}
In this work, we probed how the dyonic ModMax black hole parameters would affect the shadow, lensing, and neutrino propagation in its vicinity. To do so, we initially constrained the values of $Q_\text{m}$ under the effect of $\gamma = 1$ and $\gamma = -1$ charge screening parameters using the EHT data on the shadow diameter of Sgr. A* and M87*. Focusing on $Q_\text{e} = 0.01M - 0.10M$, since astrophysical black holes may contain a nearly zero electric charge,  we see that the value of $\gamma$ is irrelevant near $Q_\text{m} = 0$. For Sgr. A* occurs above the observed shadow diameter, while it occurs below for M87*. With these parameters, the deviation to the observed shadow radius is still within $2\sigma$ uncertainty, while for M87*, lower bounds for $Q_\text{m}$ can be observed. See Fig. \ref{sha_cons}.

We also analyzed the shadow cast behaviour for a static observer and one that is comoving with cosmic expansion for a different model of Universes (i.e., dark energy, matter, and radiation-dominated Universes). With the parameters chosen from the constraints, we saw that the shadow radius decreases due to $Q_\text{m}>0$, and such an effect is amplified as $\gamma$ decreases. As $r_\text{obs} \to \infty$, the rate of change in the shadow radius is zero. However, when the observer is comoving with the cosmic expansion, such change is no longer zero. See Fig. \ref{shacom}. While empirical data favours that our Universe is a dark energy-dominated one, it is interesting how we can also confirm this by using the deviation observed from the shadow cast. According to Fig. \ref{shacom}, such a deviation is amplified if one observes the shadow of a black hole near the cosmological horizon.

Next in our analysis is how the Modmax parameters are manifested through the deviations on the weak deflection angle, which is so sensitive even for large $b/M$. Our results revealed that the effect is amplified when $u \to 0$. Furthermore, it was shown that relativistic massive particles give a larger deflection angle than photons. As $\gamma$ decreases, so does $\hat{\alpha}$, which is also manifested in Einstein ring formation. With the chosen parameters, these deviations can be detected by current sophisticated experiments. For instance, the EHT is capable of achieving an angular resolution of $10\mu$as, which is used in mapping the stellar distributions near Sgr. A*. Furthermore, the ESA's GAIA mission can provide a sensitivity from $20\mu$as to $7\mu$as, which still depends on certain stellar magnitudes, used to map the Milky Way galaxy \cite{Liu_2017}. The smaller the angular resolution, the more the device can probe deflection at larger impact parameters. Note that the eikonal limit is independent of the spin of the perturbation, so that black holes' scalar, electromagnetic, and gravitational perturbations give the same behaviour in the eikonal limit \cite{Kodama:2003jz}. Table \ref{tab:table2} shows that the real parts decrease, but the imaginary part of the QNMs increase with the increasing parameter $\gamma$. We can conclude that these modes are stable cause the imaginary parts of the QNMs frequencies are negative. The decay rates (imaginary part) of QNMs frequencies increase with the increase of the screening parameter $\gamma$.

In Sect. \ref{sec4}, the spherically free-falling accretion model on the ModMax BH from infinity has been investigated to provide realistic visualization of the shadow cast with the accretion disk. To do so, we first calculate the flux numerically to illustrate the effects of the parameters of the ModMax $\gamma$ on the specific intensity seen by a distant observer for an infalling accretion in Figs. (\ref{fig:thinacc1}, \ref{fig:thinacc2} and \ref{fig:thinacc3}) where show the specific intensities for various values of the parameter $\gamma$ versus $b$ observed by the distant observer. We have concluded that increasing the value of $b$, increases the intensity first. Afterwards, intensity reaches the peak value sharply, where the photons are captured by a black hole quickly. There is a peak value, and then intensity slowly decreases. Furthermore, it is shown that the shadow cast of the black hole in the 2-dimensional image with a photon sphere by a distant observer in (X, Y) plane where the dark centre of it the event horizon is located, and it is circled by a bright ring with a strongest luminosity (photon sphere). It can be seen that brightness decreases gradually after the maximum region. As a result, we have concluded that the effect of the screening parameter $\gamma$ on the black hole luminosity of the shadow cast where the intensity decreases with increasing the value of the screening parameter $\gamma$ as seen in Fig. \ref{fig:thinacc3}.

In Sect. \ref{sec5}, we calculate the quasinormal modes (QNMs) of the Dyonic ModMax BHs by applying the method of WKB approximation. To study QNMs of the ModMax BHs, first, we use a massless scalar field
perturbation in the background of the black hole \ref{spherical-le}. The dependence of the QNM frequencies on the screening parameter $\gamma$ is qualitatively different for lowest and higher multipoles, as seen in Table \ref{scaltablew}. We can see in Table \ref{scaltablew} and  \ref{scaltablew2} that both real and imaginary parts of the $\omega$ decrease when the parameter $\gamma$ is increased. The scalar field QNMs for different values of $l$ is plotted in \ref{fig:QNMspectrum}, as well as for $l=0,100$ for different values of $\gamma$ are presented in Figs. \ref{WKB-different-l}, \ref{WKB-different-l100} and \ref{WKB-different-converge}. Then, we study the eikonal quasinormal modes for dyonic ModMax black holes. This method is also known as the geometric optics method due to its relation with the parameters of the null geodesics. Furthermore, the time-domain profiles of the scalar field perturbation for $l=0$ and $l=2$ cases have been plotted numerically in Figs. \ref{fig:waveforml2} and \ref{fig:waveforml3}. The ModMax black hole is given in blue; on the other hand, the Schwarzschild case is black. The logarithmic plot shows that the ModMax frequency is slightly lower. For the $l=0$ case, the plot shows the relatively short period of quasinormal ringing, making it hard to extract values of frequencies with good accuracy. After calculating the fundamental frequencies, we have provided constraints for the allowed range of masses of oscillating ModMax black holes which LIGO and LISA can detect.

Moreover, in Sect. \ref{sec6}, first we investigated the greybody factor of the ModMax black hole by
using the rigorous bound. We show the effect of the screening parameter $\gamma$ on the greybody bound for the scalar field of the ModMax black hole in Fig. \ref{fig:greybody} where it is seen that when the value of $\gamma$ parameter increases, the greybody bound $T_b$ grows larger. Second, the high-energy absorption cross-section has been studied by applying the Sinc approximation for the ModMax black hole. We have shown in Fig. \ref{fig:totabsorption} that the total absorption cross section for various values of $\gamma$ where shows that the greater values of screening parameter $\gamma$, the total absorption cross section becomes more and more unstable for higher energies and exhibits the largest amplitude. Furthermore, there is a regular oscillatory behaviour around the high-frequency limit. Fig. \ref{fig:totabsorption} presents the absorption spectrum as a function of the frequency, making clear oscillations characteristic of a diffraction pattern, where the oscillations occur around its constant value of the geometrical optics with decreasing amplitude and constant period. 

We have analyzed the neutrino oscillations and spin-flip phenomena in the dyonic ModMax black holes field. 
This analysis can be relevant in relation to the recent observations of the event horizon silhouette of a supermassive BH \cite{EventHorizonTelescope:2019dse,EventHorizonTelescope:2022xnr}.
The accretion disk around BHs is a source of photons and neutrinos \cite{Lambiase:2022ywp}. The neutrino flavour oscillations and spin oscillations may
affect the neutrino flux expected in a neutrino telescope.
We have seen that neutrino oscillation lengths get modified of $\sim 10\%$ in dyonic ModMax geometry, as compared to GR spacetime. Spin flip oscillations, instead, are suppressed, as in GR. This is a consequence of the fact that one has to consider relativistic neutrinos.
Moreover, we have only considered the effects of the gravitational field, while a complete analysis should include the magnetic field  \cite{Dvornikov:2020oay}. We have also discussed the influence of the dyonic ModMax geometry on the nucleosynthesis processes, getting a difference $\sim 5\%$, with respect to GR, which could be relevant in the Supernovae evolution. 
We conclude by noting that a follow-up of these studies will be:  the propagation of neutrino in both gravitational and electromagnetic fields; the coupling of neutrino magnetic momentum in non-linear electrodynamics;  the neutrino spin oscillations (in gravitational fields) can be potentially observed in core-collapsing SN (in our Galaxy) since a huge amount of neutrinos are emitted. These effects are expected to be probed with future neutrino telescopes. All these topics will be faced in future works.

\acknowledgements

The work of G.L. and L.M. is supported by the Italian Istituto Nazionale di Fisica Nucleare (INFN) through the ``QGSKY'' project and by Ministero dell'Istruzione, Universit\`a e Ricerca (MIUR). G.L., A. {\"O}. and R. P. would like to acknowledge networking support by the COST Action CA18108 - Quantum gravity phenomenology in the multi-messenger approach (QG-MM).

\bibliography{references}
\bibliographystyle{apsrev}

\end{document}